\documentclass[fleqn,usenatbib]{mnras}

\usepackage{newtxtext,newtxmath}

\usepackage[T1]{fontenc}

\DeclareRobustCommand{\VAN}[3]{#2}
\let\VANthebibliography\thebibliography
\def\thebibliography{\DeclareRobustCommand{\VAN}[3]{##3}\VANthebibliography}

\usepackage{graphicx}	
\usepackage{amsmath}	
\usepackage{booktabs}
\usepackage[switch, modulo]{lineno}
\usepackage{float}
\usepackage{comment}
\usepackage{xspace}
\usepackage{verbatim}
\usepackage{longtable}

\newcommand\rp{\ensuremath{r^\prime}\xspace}

\title[A Targeted Search for Lensed Supernovae]{A targeted search for strongly lensed supernovae with the Las Cumbres Observatory}

\author[P. Craig et al.]{
Peter Craig,$^{1}$\thanks{E-mail: craigpe1@msu.edu}
Kyle O'Connor,$^{2}$
Sukanya Chakrabarti$^{3}$
Steven A.\ Rodney$^{2}$
Justin R. Pierel$^{4}$
\newauthor
Curtis McCully$^{6}$
and Ismael Perez-Fournon$^{7,8}$
\\
$^{1}$Center for Data Intensive and Time Domain Astronomy, Department of Physics and Astronomy, Michigan State University, East Lansing, MI 48824, USA\\
$^{2}$University of South Carolina, 712 Main St., Columbia, SC 29208, USA\\
$^{3}$Department of Physics and Astronomy, University of Alabama, Huntsville, Huntsville, Alabama 35899\\
$^{4}$Space Telescope Science Institute, Baltimore, MD 21218\\
$^{6}$Las Cumbres Observatory, 6740 Cortona Dr Ste 102, Goleta, CA 93117-5575, USA\\
$^{7}$Instituto de Astrof\'\i sica de Canarias, C/V\'\i a L\'actea, s/n, E-38205 San Crist\'obal de La Laguna, Tenerife, Spain\\
$^{8}$Universidad de La Laguna, Dpto. Astrof\'\i sica, E-38206 San Crist\'obal de La Laguna, Tenerife, Spain\\
}
\date{Accepted XXX. Received YYY; in original form ZZZ}

\pubyear{2024}

\begin{document}
\label{firstpage}
\pagerange{\pageref{firstpage}--\pageref{lastpage}}
\maketitle

\begin{abstract}

Gravitationally lensed supernovae (glSNe) are of interest for time delay cosmology and SN physics. However, glSNe detections are rare, owing to the intrinsic rarity of SN explosions, the necessity of alignment with a foreground lens, and the relatively short window of detectability. We present the Las Cumbres Observatory Lensed Supernova Search, LCOLSS, a targeted survey designed for detecting glSNe in known strong lensing systems. Using cadenced \rp-band imaging, LCOLSS targeted 114 galaxy-galaxy lensing systems with high expected SN rates, based on estimated star formation rates. No plausible glSN was detected by LCOLSS during our two year observing program.  We carry out an analysis here to measure a detection efficiency for these observations.  We then perform Monte Carlo simulations using the predicted supernova rates to determine the expected number of glSN detections. The results of the simulation suggest an expected number of detections and $68\%$ Poisson confidence intervals, $N_{SN} = 0.20, [0,2.1] $, $N_{Ia} = 0.08, [0,2.0]$, $N_{CC} = 0.12, [0,2.0]$, for all SNe, Type Ia SNe, and core-collapse (CC) SNe respectively. These results are broadly consistent with the absence of a detection in our survey. Analysis of the survey strategy can provide insights for future efforts to develop targeted glSN discovery programs, especially considering the large anticipated yields of upcoming surveys.
\end{abstract}




\section{Introduction} \label{sec:intro}
Supernovae (SNe) are violent stellar explosions caused by dying stars. SNe are often utilized as cosmological probes due to a combination of large luminosities and predictable light curves. Type Ia SNe in particular are commonly used to measure distances to other galaxies, as their luminosity can be accurately predicted from their light curves \citep{phillips_absolute_1993}. Using Type Ia SNe as standard candles in the local universe has allowed for accurate measurements of the Hubble Constant (H$_0$) to be made based on redshifts combined with distances derived from SNe \citep{Zhang2017,Riess2019,Wong2020,Riess2021,Riess2022}. Distance measurements made with SNe allowed for measurements of the accelerating expansion of the universe for the first time \citep{riess_observational_1998, perlmutter_constraining_1999}. New measurements of the H$_0$ remain in high demand despite the accurate SNe measurements, as comparably accurate measurements utilizing the cosmic microwave background (CMB) produce a significantly different result. The tension between the results from these methods has now reached the $5 \sigma$ level \citep{Wong2020,Riess2022,Planck2020}.

SNe of all types can provide cosmological constraints when multiply-imaged by strong gravitational lensing \citep{refsdal_gravitational_1964}. When aligned with a massive foreground object, the light from the SN will be deflected as it passes through the gravitational potential. In the strong lensing regime the background object may be observed as multiple distinct images on the sky, with relative time delays between these images that can range from hours to years. Measurement of the relative time delays, combined with a model of the lensing potential, allows for a measurement of $H_0$ and other cosmological parameters \citep{Kelly2022}. Lensed quasars have been used in this way for many years \citep{Suyu2018}. SN time-delays are attractive additions to the cosmologist's toolkit, owing primarily to the predictability of their light curves. They also allow for measurements with shorter observing timescales in each image and can provide Type Ia SN luminosity distance measurements to break the mass-sheet degeneracy \citep{Oguri2003}.

To date, only 7 strongly lensed supernovae (glSNe) with multiple images have been detected \citep{Kelly2016,Goobar2017,Rodney2021,Kelly2022,Goobar2022,Chen2022,Frye2023}. The most recent detection, SN H0pe, has now been used to make a measurement of H$_0$ that takes advantage of the standardizable Type Ia luminosity \citep{Chen2024, Pascale2024,Pierel2024}. However, this sample is expected to grow dramatically with discoveries from new transient surveys using the Vera Rubin Observatory and the Roman Space Telescope \citep{goldstein_how_2016,wojtak_magnified_2019,Pierel2021}. These high-cadence, wide-field surveys may detect hundreds of glSNe over the next decade, though the majority will have relatively short time delays and small angular separations \citep{Pierel2021,Huber2019}.

One approach for hunting glSNe is to monitor a sample of known strong lensing systems, which is the strategy that we adopted with our survey, the Las Cumbres Observatory Lensed Supernova Search (LCOLSS). With the release of the SLACS sample \citep{Bolton2008}, the SWELLS sample of lensed spiral galaxies \citep{Treu2011,Brewer2012}, the SLACS for the Masses Survey (S4TM) \citep{Shu2017}, and the BELLS sample \citep{Brownstein2012}, we have a large enough number of known strong lensing systems at low redshift ($z < 1$) to plausibly make a detection over $\sim$ year timescales in a targeted search. Most of these galaxies have been presented in \citet{shu_prediction_2018} (hereafter S18), which examined a sample of 128 galaxy-galaxy strong lensing systems (i.e., the sources in SLACS, S4TM and BELLS, 98 of which are included in our sample) in which the lens is an early type galaxy and the background galaxy has strong star formation. The high star formation rates (SFRs) of the lensed galaxies should result in a higher rate of SN explosions (including Type Ia SNe), and the lensing galaxies with large Einstein radii would result in longer observable time delays with a relatively simple lensing mass profiles \citep{Li2011,Oguri2019}. The SFRs and the SN rates are discussed in more detail in S18. This paper will present the (null) results of our pilot glSNe search program. The goal of LCOLSS was to detect a glSN at low redshift (our sample has an average source redshift of 0.66), and to provide a measurement of the H$_0$ from time delays. The time delay is primarily sensitive to H$_0$, in contrast to unlensed supernovae that use relative luminosity distances with other sensitivities \citep{Suyu2017}.

Spiral galaxies have more complex lensing potentials, making H$_0$ measurements less accurate in general for these sources due to higher uncertainties on their potentials. Our inclusion of the SWELLS sample of lensed spirals instead of only considering elliptical galaxies was motivated by the goal of determining two \emph{independent} constraints on the dark matter distribution from strong lensing and from analysis of HI maps \citep{ChakrabartiBlitz2009,ChakrabartiBlitz2011,Chakrabartietal2011,Chakrabarti2013}, which is only possible at low-redshift where HI maps can be obtained. To date, there is only one detection of HI in emission in a known strong spiral lens \citep{Lipnickyetal2018}, though we may expect detections and HI maps of strong spiral lenses from next-generation HI surveys with the Square Kilometer Array (SKA) \citep{Catinella2015,Giovanelli2015}. The highest redshift HI map thus far from an unlensed starburst galaxy was obtained by \cite{Fernandezetal2018} at $z = 0.376$.

\section{LCOLSS Survey} \label{sec:survey}

The LCOLSS survey ran for two years, beginning in late 2018 and ending in the middle of 2020. During this window of operation, regular images were taken for a sample of lensing systems at a typical cadence of two weeks, always in groups of at least two images. This cadence is high enough to provide decent chances of detecting a glSN in one of our survey targets. Each set of observations were reduced manually through difference imaging pipelines to search for supernovae (and in some cases other transients) in these images, when compared against data from earlier in the survey. This section will describe the details of our lensing systems, and the observations taken of these sources.

\subsection{Lensing Systems}\label{sec:lens_sys}

Our sample includes 114 galaxy-galaxy strong lensing systems. A table identifying these systems and containing relevant lensing information, such as magnifications and SN rates, can be found in Appendix \ref{app:lens_data}. In S18 the SFR estimates were derived from emission line flux density measurements, but an error in the calculation resulted in an {\it overestimation} of the inferred SFRs, proportional to $(1+z)^2$ \citep{Shu2021erratum}. We have corrected for this error and propagated it through to also correct the inferred Type Ia and Core Collapse (CC) SN rates, including all the subtypes of CC SNe. The net result is a change of -5.53 SNe yr$^{-1}$ in the expected total SN explosion rate, summed over the 98 S18 sources in our sample. Appendix \ref{app:lens_data} also includes information about the number of visits to each source, along with the average seeing and airmass for our observations.

The targets for LCOLSS were selected from a combination of the SLACS, S4TM, BELLS and SWELLS catalogs. We begin with the 128 sources listed in S18, which are from a combination of SLACS, S4TM and BELLS. All SLACS and S4TM sources in that paper that include estimates on the SN rates are included in the sample, making up the majority of our sources. This provides 55 SLACS sources and 34 S4TM sources. An additional 9 BELLS sources are included in our sample, yielding a total of 98 sources that were analyzed in S18. The BELLS sources are typically at higher source redshifts than the SLACS sources. The included BELLS sources are a subset of the BELLS sources included in S18, which is because the BELLS sample is at higher redshift, where detecting a glSN would likely be more difficult, except in cases with large lensing magnifications.

In addition to these sources, we include the SWELLS sample of spiral lenses. There are 20 SWELLS sources that are Grade A lenses (as defined by the SLACS survey \cite{Bolton2006,Bolton2008}). Of these, 4 are already included in the sample from the SLACS sample. The remaining 16 sources are added to our catalog, leaving a total sample size of 114 lenses.

\begin{figure}
    \centering
    \includegraphics[width=0.95\columnwidth]{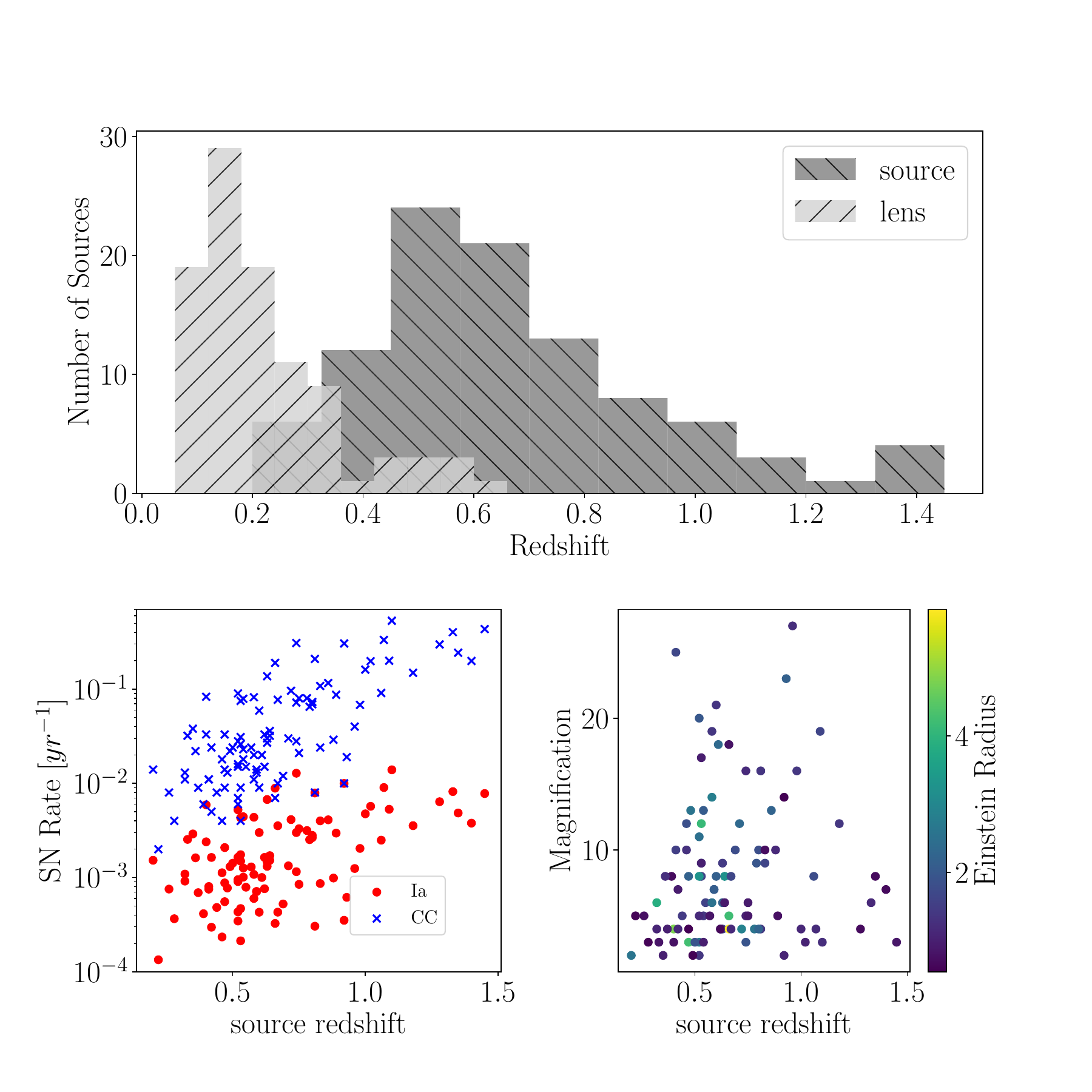}
    \caption{The top panel shows the redshift distribution for our 98 S18 targets, with lensing galaxies in light grey and lensed background galaxies in dark grey. The bottom left panel shows the predicted (observer frame) SN rates for each of the lensed source galaxies, plotted against redshift. Each of the lensed source galaxies appears twice: in red for the Type Ia SN rate and in blue for the CC SN rate. The bottom right panel shows lensing magnifications of each of the systems, plotted against source redshift. The colorbar in this panel shows the Einstein radii for our sources. See Appendix Table \ref{Atab:yiping_data} for these values, and \citep{shu_prediction_2018} for details of the redshift measurements, SN rate estimates, and lensing magnifications. Einstein radii are quoted in the original survey papers \citep{Bolton2008,Brownstein2012,Shu2017}. Note that the supernova rates here have been corrected for the error in S18 as discussed in section \ref{sec:lens_sys}. }
    \label{fig:target_data}
\end{figure}

The SN rates are determined in S18 from the SFRs of the background source galaxies. The SFRs are measured from observed SDSS and BOSS spectra, using integrated fluxes from [OII] 3727{\AA} emission doublet, $F_{OII}^{\rm{intrinsic}} = F_{OII}/\mu/f_{\rm{fiber}}$. Determining intrinsic flux in the source galaxy is a multistep process. The measured flux in the source galaxy, $F_{OII}$, is extracted from the spectrum centered on the lensing galaxy, with the lensing galaxy spectra subtracted, and a double-Gaussian plus line is fit to the [OII] doublet of the remaining source galaxy. The fraction of the source galaxy captured in the spectrograph, $f_{\rm{fiber}}$, is then determined using a convolution of the lensed HST image with the PSF from the ground based spectrograph. Finally, this value is demagnified to give the intrinsic flux. The intrinsic flux with [OII] then is calibrated to a SFR, using a subsample for which H$\alpha$ has been detected, as the H$\alpha$ luminosity scales directly with ionizing flux. The uncertainties on the rates based on the flux measurement are large due to the lensing galaxy subtraction, de-magnification, and fiber loss. There are also sources for biases; perhaps the largest being in the inability to perform a system by system H$\alpha$ calibration for the SFRs; i.e. even before the final models to convert the star formation into SN rates. Another potential source of bias is on the relative CC rates which were based on local measurements and assumed to have no redshift evolution, which remains an open question in general \citet{strolger_delay_2020}. The typical uncertainties on the derived SN rates are around 57$\%$, with the smallest uncertainties at $45\%$. This applies to both the rates for Type Ia SNe and for CC SNe. The average percentage uncertainties are $56.7\%$ and $56.8\%$ on the Type Ia rates and CC rates, respectively \citep{shu_prediction_2018}.

\begin{figure}
    \centering
    \includegraphics[width=0.95\columnwidth]{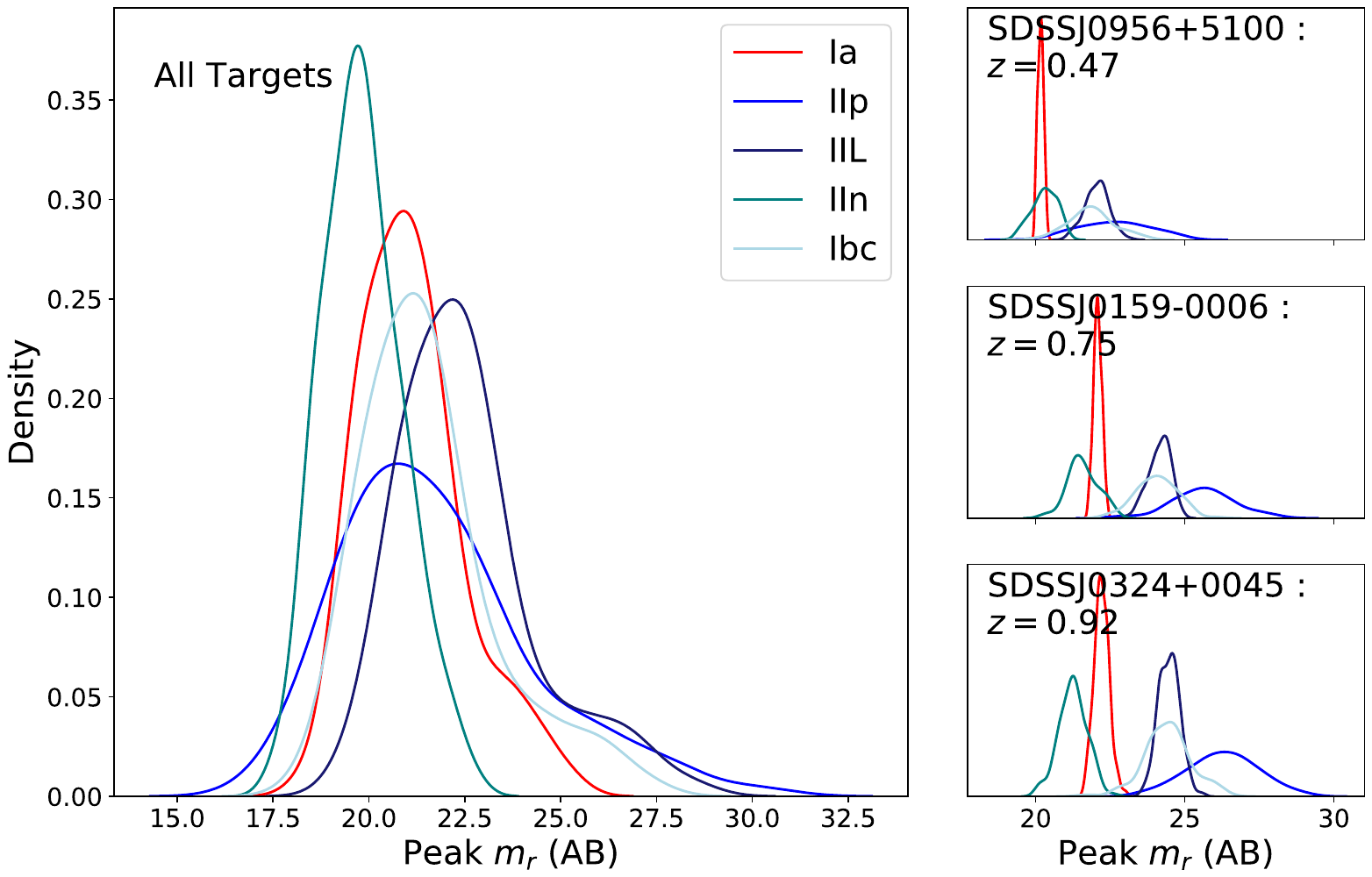}
    \caption{The distributions of peak r-band magnitudes for each SN subtype, including the lensing magnifications and dust effects. On the left is the distribution corresponding to all of our targets that were included in S18. On the right are a few examples for individual targets at different source redshifts.}
    \label{fig:SNpks}
\end{figure}

For more details on the redshift measurements, SFR calculations, SN rate estimates, and lensing magnifications, see \citet{shu_prediction_2018} and \citet{shu_sloan_2017}. The distributions of SN rates, redshifts (both for source and lens galaxies), and the magnifications can be seen in Figure \ref{fig:target_data}. The resulting distributions of the peak brightness, in the \rp filter, for the various SN types are shown in Figure \ref{fig:SNpks}. This figure also displays the peak brightness distributions for three individual sources.

\subsection{Observations}
\begin{figure*}
    \centering
    \includegraphics[width=0.95\textwidth]{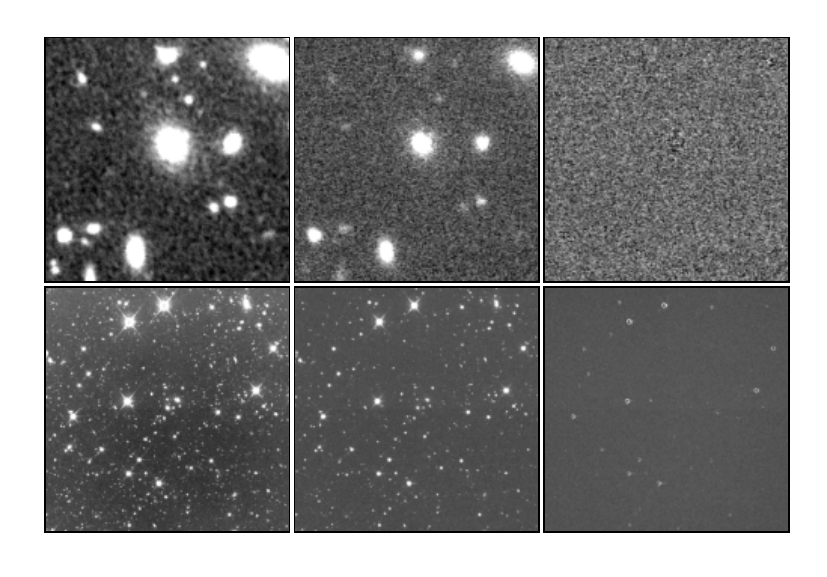}
    \caption{An example of a \textsc{pydia} generated difference image. The left column shows the template, the middle column shows the search image, and the right column shows the difference image. The top row shows a cutout around the target, while the bottom row shows the entire image. The artifacts in the bottom right panel are remnants of bright stars being subtracted from themselves, and are not transient sources.}
    \label{fig:pydia_ex}
\end{figure*}

The Las Cumbres Observatory (LCO) is a robotic telescope network designed for time-domain astronomy \citep{Brown2013}. For LCOLSS we used the 1-meter LCO telescopes and Sinistro cameras, typically collecting three 300s exposures per target through the \rp filter. Our observations reached 3$\sigma$ limiting magnitudes ranging from $\sim$21-23 AB mags. We generated difference images using the software tools \textsc{pydia} \citep{albrow_pydia_2017} and \textsc{hotpants} \citep{becker_hotpants_2013}. Each difference image was visually inspected for potential transient sources by at least two team members. An example difference image generated by \textsc{pydia} can be seen in Figure \ref{fig:pydia_ex}. Our photometric accuracy was confirmed by comparing our photometry with SDSS magnitudes, typically providing errors of $\sim 0.05$ magnitudes. An example comparison with SDSS magnitudes can be seen in Appendix \ref{app:lens_data}. We derived zero points for each image by matching detected stars to SDSS, and limited to the magnitude interval $\rp = [16,20]$ to ensure high signal-to-noise and no saturation effects.

Our exposure times varied based on our rate of time usage (relative to the time award in a given semester), typically either increasing the exposure time to 600 seconds per frame or reducing the number of frames by one. We calibrated our image processing pipeline by planting fake sources into the images and measuring the recovery efficiency as a function of magnitude. More information about the distribution of limiting magnitudes for our survey may be found in Appendix \ref{app:lens_data}, based on the predictions of the  LCO Exposure Time Calculator\footnote{
 \url{https://exposure-time-calculator.lco.global/}} (ETC).

The template images used for this survey were generated from images taken earlier in the survey. The initial templates were the first images of each source obtained during the survey. As the survey progressed, new templates were generated by stacking all prior observations (excluding observations with poor data quality, bad weather, or other problems). These stacked templates provided deeper template images available for the rest of the survey. After approximately 1 year, the stacked templates were deep enough that further stacking provided minimal benefits, so new templates were only generated after this for sources that had low quality templates due to lower numbers of suitable observations.

In general, we focused our search on a region of approximately one arcminute around the target galaxy. The area within $\sim 10''$ of the target galaxy (which for most sources is more than 5 times the Einstein radius) is given particular attention, as this is likely where a glSN would appear. In this region, any possible SN will be followed up, since our expected yield is low it is important not to miss any potential detection. This helps to mitigate the risks of misclassifying a glSN as a SN in a foreground galaxy. If such a SN was detected, it would be reported and followed up spectroscopically, at which point we would be able to confirm the source's origin as a foreground SN. As a result, contamination from nearby transients would not negatively impact our detection rates (which is critical for a survey of this nature where the expected number of detections is of order 1).

Out to one arcminute, we still search carefully for potential candidates, to cover any glSNe that might have formed in the outskirts of the source galaxy. As we get farther from the target, for the sake of time, the manual search in the difference images becomes less thorough, although clear transients, primarily solar system objects, in this region were still often identified. Transients close to our detection threshold, but far from the target system, are more likely to be missed or to lack follow-up observations. The detection of solar system objects in our difference images provides confirmation that transients were successfully recovered, but because most solar system objects would appear at significant distances from the target source, the recovered objects were typically significantly brighter than our limiting magnitude.

\section{Methods} \label{sec:methods}

In this section we describe the methods used to estimate the glSN yield for LCOLSS and similar surveys. The method operates by planting simulated SNe into our images, then determining how likely we would be to recover those sources. The recovery chance is determined by measuring a detection efficiency, defined as the magnitude dependent ratio between the number of detected sources and the total number of sources, for a given image. This is done for many simulated SNe in a Monte Carlo simulation, which is ultimately responsible for the yield estimate of our survey. We begin in section \ref{sec:plant}, which describes a method for planting simulated SNe into our difference images. A discussion of our detection efficiencies can be found in section \ref{sec:det_eff}. Finally, we discuss our Monte Carlo simulations in section \ref{sec:MC}

For the analysis in this paper, we focus on the 98 sources contained in S18. These sources have readily available measurements for the SN rates, lensing magnifications, lens models and redshifts. Also, by restricting the analysis to this sample, we ensure that the lens modelling techniques are consistent across the sample. As a result of this, the detection rate estimates calculated below only include 98 of our 114 sources, but we expect that the change in the results caused by the other sources in our sample should be small, and will be less than the uncertainties on our yield estimates. For our survey, this will produce a somewhat conservative detection rate, but should produce a reasonable estimate of the typical detection rate per source. 

\subsection{Planting Simulated SNe}\label{sec:plant}

The first step in planting a simulated SN is to measure the zero point (ZP) for each image such that fake sources can be realistically planted at different magnitudes. Stars in each LCO image field are matched to SDSS data which have the stellar magnitudes in the identical \rp filter. Fluxes for the stars on the LCO images are then measured with both an aperture-photometry and PSF-photometry method. The zero points of the data images are determined using the least squares fit to $m_{sdss} = -2.5\log_{10}{f_{lco}}$; where the calibration is restricted to those stars inside the interval $\rp = [16,20]$. This ensures that the calibration stars have a good signal-to-noise and no saturation. Our photmetric accuracy is typically better than 0.1 magnitude, and a sample distribution used for zero point calibrations can be seen in Appendix \ref{app:lens_data}.

A gaussian based model, \textit{IntegratedGaussianPRF}, is used for the PSF,  available in the \texttt{photutils} software. The spread for the model is taken from the full width at half maximum value, \texttt{L1FWHM}\footnote{\texttt{L1FWHM} is defined by the LCO pipeline as the frame FWHM calculated from the SExtractor detections with very small objects and those with FLAGS != 0 weeded out. The individual FWHM values are calculated slightly differently than the SExtractor FWHM values, using FWHM = $2 * \sqrt{ln(2) (a^2 + b^2)}$, where a and b are the source extractor parameters for the maximum and minimum spatial RMS. L1FWHM is then a clipped and filtered average over all suitable sources in the frame.}, determined in the LCO pipeline which is available in each image's header \citep{mccully_real-time_2018}. The background around each known star is determined and subtracted using a localized sigma-clipped standard deviation on image pixels. The flux is then fit for using the amplitude of the model with fixed positions on the background subtracted sources. The aperture method for the flux uses a circle with radius, $r = 2 \texttt{L1FWHM}$, again with a local background subtraction, taken from the sigma-clipped standard deviation on pixels in an annulus extending over $r_{ann} = [r+5,r+10]$.

The zero point is determined for a large sample ($ \gtrsim 700$) of data images from our survey. This sample covers the typical range of image properties found across our survey, exploring the full range of airmasses, exposure time and various moon related parameters. In order to reduce computational time on the large number of remaining images, we set up a nearest neighbor interpolator to estimate the zero point from an image header. The accuracy that can be expected in predicting a zero point is determined using a nearest neighbor interpolation on the data, over image header parameters. The relevant parameters are the \texttt{EXPTIME}, \texttt{MOONFRAC}, \texttt{MOONALT}, \texttt{MOONDIST}, and \texttt{AIRMASS}. The difference between the measured zero point and this interpolation method gives a distribution of errors, $\Delta ZP = ZP - ZP_{interp}$, is used to estimate the accuracy of obtaining a zero point from header parameters. This gives us an accuracy of $<\Delta ZP> = 0.17$ mag across our testing sample. This provides a computationally inexpensive method to estimate the ZP for many images, and one that can be fully automated with ease.

After the zero point calibration, the PSF model is scaled to different magnitudes and added to the difference image as simulated SNe. These SNe represent singular images of a SN, and don't account for multiple lensed images that could be blended and unresolved. Images with simulated SNe can now be used for measuring detection efficiencies and estimating our survey yield. These procedures take advantage of various \texttt{ photutils} functions. The PSF used, as mentioned earlier, is the \textit{IntegratedGaussianPRF}, which requires only the spread taken from the image headers. The planting of the PSF in the image at various positions and fluxes is done using \textit{subtract\_psf}. The fluxes of the synthetic SNe that we have planted are determined using the zero point, at magnitudes covering $r = [20.5,23.5]$. We expect that all sources at the bright end of this range will be recovered, and nearly 0 sources at the faint end should be recovered.

\subsection{Measuring the Detection Efficiency}\label{sec:det_eff}

We determine the detection efficiencies in our survey following the methods of \citep{strolger_rate_2015}. A simple efficiency curve is a step function, where any SN brighter than a specified magnitude is detected. For instance, in our survey such an assumption would be that for our deeper images any supernova with an apparent magnitude of 23 or brighter is detected. This is our predicted limiting magnitude based on ETC calculations for our typical observations, so in principal sources at this magnitude are detectable. This assumes that they remain sufficiently visible in our difference images, which is not a guarantee. Additional effects, like noise, weather, and moonlight, can also reduce our limiting magnitude, adding the potential of missing some sources close to this magnitude threshold.

We apply here a more robust detection efficiency by planting synthetic point sources (as described above) into the difference images and subsequently recovering them in the images. Following \citet{strolger_rate_2015}, the fractional detection efficiency is then fit as an efficiency parametric model with an exponential functional form. The model parameters, $m50$ and $\alpha$, characterize the magnitude at which an object has a fifty percent chance of being detected, and the steepness of the efficiency fall off, respectively. In other words, we write the fractional detection efficiency, $\epsilon$, as:
\begin{equation}
\epsilon = \big( 1+e^{\alpha(m-m50)} \big)^{-1}  
\label{eq:eff}
\end{equation}

where $\epsilon$ is the efficiency and m is the apparent magnitude of a source. We chose to be more rigorous with the efficiency method here since the expected yield is low, and LCO has not been used for a targeted glSN survey like this before. 

The synthetic SNe for detection efficiency measurements are planted in a twenty-five position lattice configuration around the lensing galaxy. These sources are separated by 25 pixels, which covers a total area of 0.42 square arcminutes, with a scale of $0.389''$ per pixel. At the redshifts of our sample, this covers a physical distance of tens to hundreds of kpc, and could represent any possible SNe within the source galaxy, including SNe that form in the outskirts of galaxies \citep{Chakrabarti2018}. The lattice is centered on the target, and does not vary based on source properties. In determining the recovery for the fake planted sources, we have used \textit{DAOStarFinder} with a $3 \sigma -$ threshold, where $\sigma$ is the difference image noise taken as the sigma-clipped standard deviation. This method looks for sources that have at least 5 adjacent pixels above the given threshold. A linear least squares method is applied to the recovered fractions at the given magnitudes to determine the parameters $m_{50}$ and $\alpha$ in the efficiency function, Equation \ref{eq:eff}. 

\begin{figure*}
    \centering
    \includegraphics[width=0.95\textwidth]{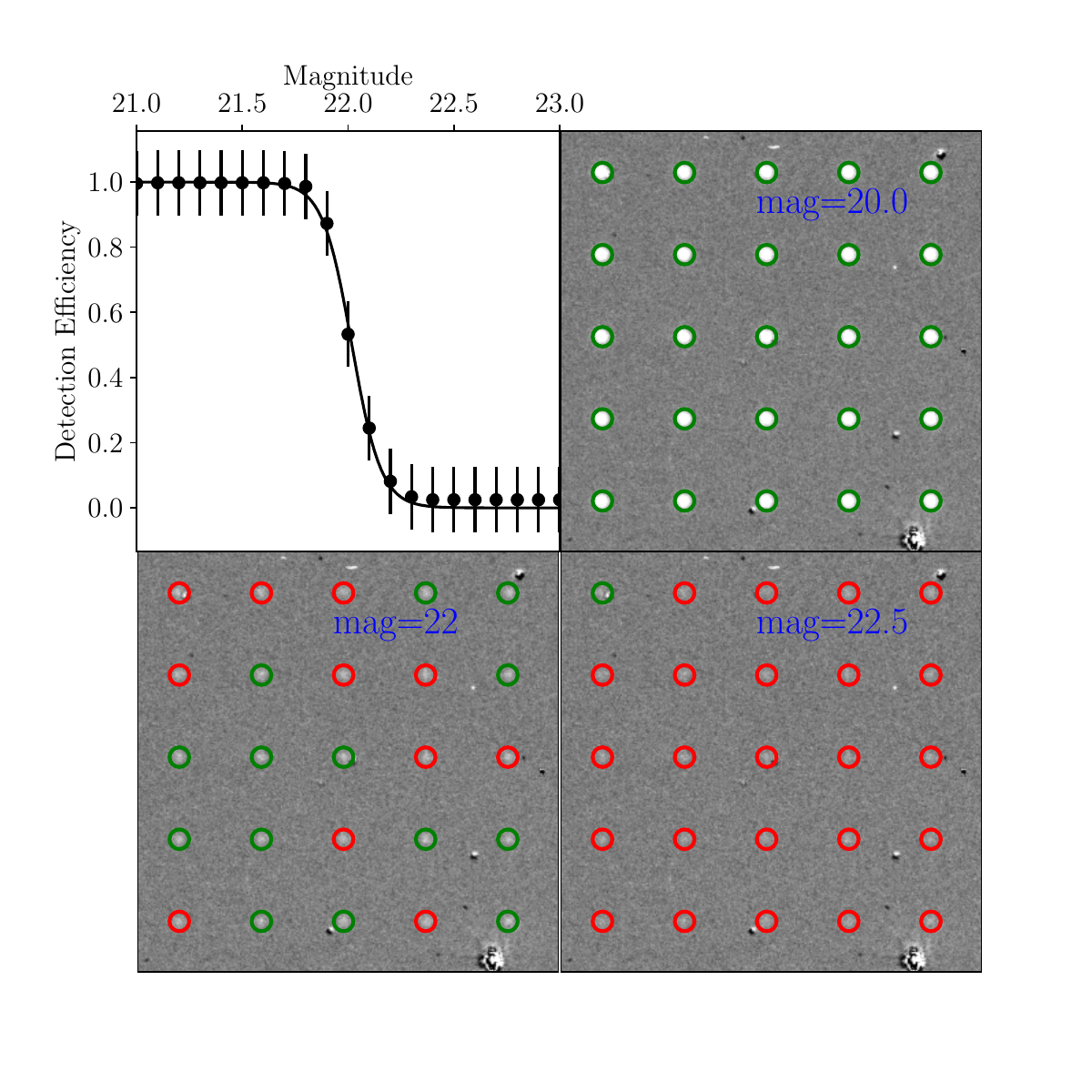}
    \caption{An example of the efficiency measurement is shown above. The top left figure shows both the measured detection efficiencies and our best fitting curve from Equation \ref{eq:eff}. The top right image has plants with magnitudes which would provide a near certain detection. On the bottom left is an image with plants at magnitude near the fifty percent efficiency value. The bottom right has plants with magnitudes faint enough that detection would be unlikely. Green and red circles mark the locations of the plants and signify detected and missed objects respectively.}
    \label{fig:fit_ex}
\end{figure*}

An example showing the planting of SNe into a difference image and the measured efficiency are shown in Figure \ref{fig:fit_ex}. This is for one data frame that has a measured $m50$ of 22.01, and shows the measured efficiency as a function of magnitude. The case with a source magnitude of 22 is shown in the bottom left, where close to half of the synthetic sources are detected. The source detection software was set to agree with those objects which would have likely received close attention; which can be seen in Figure \ref{fig:fit_ex}. Human searchers are expected to outperform this detection efficiency algorithm, so the resulting efficiency curves will be somewhat conservative.

We measure the detection efficiency across the same group ($ \gtrsim 700$) of data images used for the zero points. These results are used (in a nearest neighbor interpolation) to estimate the detection efficiency for all surveyed images ($ \gtrsim 6000$) in the four LCOLSS semesters over which we carried out our observations. This interpolation is carried out based on a variety of header parameters in our images. This includes the exposure time (\texttt{EXPTIME}), airmass (\texttt{AIRMASS}), moon altitude (\texttt{MOONALT}), moon fraction (\texttt{MOONFRAC}), and the distance between the source and the moon (\texttt{MOONDIST}), all of which are related to the detection efficiency. By only using a subsample of our set of images for this analysis we significantly reduce the computational costs, and allow for easy prediction of the detection efficiencies in other images. The sources selected were chosen randomly, and sample approximately the same distributions of these header parameters of the entire data set. Testing indicates that this interpolator is successful at predicting the $m50$ and $\alpha$ values, producing reasonable detection efficiency curves.

\begin{figure}
    \centering
    \includegraphics[width=0.95\columnwidth]{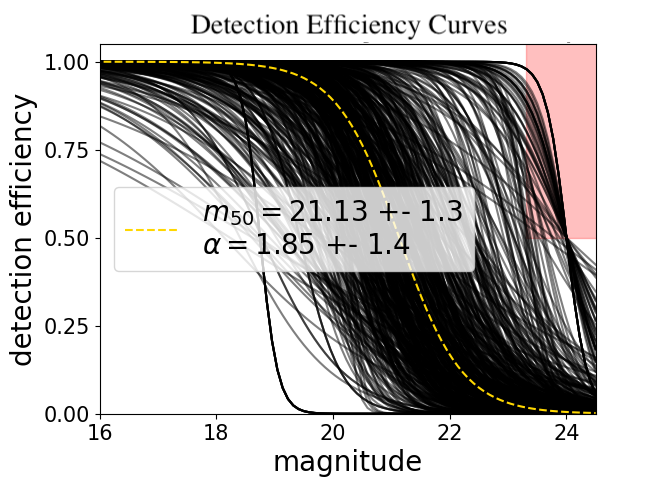}
    \caption[Measured Detection Efficiency for our Difference Images]{The curves corresponding to the fitted $m50$ and $\alpha$ parameters for Equation \ref{eq:eff} on the 572 measured data frames. The dashed gold curve shows the median $m_{50}$ and $\alpha$ curve. The red region highlights those curves which have not been used to assign efficiencies (our observations are unlikely to have detections at those depths), and a conservative value is assigned instead. The median values across this sample are $m_{50} = 21.13 \pm 1.3$ and $\alpha = 1.85 \pm 1.4$. This gives a lower $m50$ than what would generally be predicted by the ETC, but is a more realistic representation of our ability to detect sources after the images have gone through the difference imaging pipelines.}
    \label{fig:effcurves}
\end{figure}

The resulting efficiency curves, corresponding to fractions of recovered sources at different magnitudes, are shown for the fitted image data in Figure \ref{fig:effcurves}. The depth characterized by $m_{50}$ is in general agreement with the predictions made by the exposure time calculator. These observations typically detect transients around $r \approx 21$, but the depth rarely reaches much past $r \approx 23$. There is a small shift to lower limiting magnitude relative to the ETC calculations. The steepness of the efficiency drop-off, $\alpha$, indicates that a point source limiting magnitude in a given observation drops from a likely to unlikely detection on the order of $\sim 0.5$ mag.

The measured $m_{50}$ and $\alpha$ detection efficiency values are used along with image header parameters \texttt{EXPTIME}, \texttt{MOONFRAC}, \texttt{MOONALT}, \texttt{MOONDIST}, \texttt{AIRMASS}, \texttt{L1MEAN}, \texttt{L1FWHM} to establish our nearest neighbor interpolation for predicting the efficiencies in other images. The distribution of errors from these, $\Delta m_{50} = m_{50} - m_{50_interp}$ and $\Delta \alpha = \alpha - \alpha_{interp}$, provide confidence in our ability to assign efficiencies to the full set of surveyed data in the two years of observations. Based on a sample of $ \gtrsim 700$ observations, the interpolation derived $m_{50}$ values typically agree to within 0.5 magnitudes, and perhaps more importantly show no systematic offset relative to the measured detection efficiency values. In any given observation this may alter the yields slightly, but integrated across hundreds of observations this will average out, resulting in a good yield estimate.

These results in principle could be impacted by the occasional case with a poor subtraction. In most of these cases, the poor subtractions result from low data quality in the data frames (such as large amounts of sky noise or poor seeing leading to a large FWHM for the psf). These cases will be modelled by our methods from the psf modelling and zero-point calculations, resulting in reduced chances of detecting the planted sources (or real ones). Other cases can be caused by low quality template images or by something going wrong in the difference imaging pipeline. Poor template issues are primarily restricted to the first few months of data, where stacked templates over many observations were not yet available. A few cases came up where a poor subtraction was generated due to noisy observations being included in the template images, but these were quickly resolved by generating new templates that excluded the problematic observations.

Pipeline issues were occasionally problematic, with the most common issues stemming from poor alignment of the images before subtraction. Cases that one algorithm struggles with can often be resolved by using another code. Some cases of poor subtractions do remain, but represent a small fraction of the remaining images, and are not specifically accounted for in our models. Since each observation has multiple data frames, sometimes an adequate difference image can be generated from one of the frames. Creating a stacked image of these frames then leads to poor subtractions, but often we can search the one good difference image, resulting in a somewhat reduced detection efficiency. This is uncommon however, and is expected to have a minimal impact on our resulting yields. The only sources in the survey that tend to consistently have poor subtractions are cases with very few observations, where the template images are lower quality. There are not many such sources (2 sources with fewer than 5 observations, as can be seen in Appendix \ref{app:lens_data}.

\subsection{SN Monte Carlo Simulation}\label{sec:MC}

The population of lensed SNe that could be discovered by LCOLSS was determined by using a Monte Carlo method to generate supernova light curves in each target host galaxy. In this calculation, we sample over SN rates, SN spectral-temporal models, the SN intrinsic luminosity, and lensing galaxy magnification. We also include the effects of dust extinction, which is estimated based on the position of the source on the sky. 

The SN rates, $R_{SN}$, determined by \citet{shu_prediction_2018}, shown in Figure \ref{fig:target_data}, were scaled up, by a factor of $10^6$, so that many thousands of Type Ia and CC SNe were made for each target. These SN rates are the ones listed in appendix \ref{app:lens_data}. The survey's time window, $t_{lcolss}$, spanned 789 days (2.16 years), from the first observation on 2018-12-04 to the final observation on 2021-01-31. The total number of SNe generated for each target $N_{SN} = R_{SN} t_{lcolss}$ are distributed uniformly over the surveyed window. The result is that each source contains multiple SNe with explosion dates ranging across the entire survey window. Each generated SN can then be analyzed using the actual observations around the time of explosion, where our interpolators on the detection efficiency can be used to determine the probability that the SN would have been detected in those images. In each case the magnitude of the source is determined based on a light curve that is evaluated at the times of the nearby observations, such that the true cadencing of our survey is captured in this estimate. These light curves are based on randomly sampled parameters for SN light curves, to cover the range of potential SN behavior.

Light curves were generated for all the synthetic SNe with each target redshift and magnification using \texttt{SNCosmo} \citep{barbary_sncosmo_2014}; a python package with an available library of spectral-temporal models for various SNe types, and which takes into account K correction (conversion to rest frame) and the distance modulus. The SN templates, and normally distributed luminosity functions were selected following \citep{Goldstein2019}. The only difference in choice being for the Type Ia template, where \texttt{ salt2-extended} \citep{guy_salt2_2007} \citep{pierel_extending_2018}, was used. Table \ref{tab:LuminosityandTemplates} provides details on the selected templates for the SNe, along with the parameters of the peak magnitude distribution for different types of SNe.

Dust extinction was included using the model of \citet{cardelli_relationship_1989} (as another built-in feature of \texttt{SNCosmo}) for both the MW and the host galaxy. In the MW, $E(B-V)$ was assigned using the target coordinates in the map from \citet{schlegel_maps_1998}, and $R(V) = 3.1$ following a typical value determined for the diffuse interstellar medium in \citet{cardelli_relationship_1989}. With the host galaxies, the dust extinction slope, $R(V) = 3.1$, was used again. A conservative value $A_V = 0.1$ was used as each host's total extinction. This gives us a total $A_{\rp}$ ranging from 0.15 to 0.59, depending on the MW extinction.

Stretch and color parameters, $x1,c$ are required to specify the \texttt{salt2-extended} Type Ia light curves. They characterize the intrinsic Type Ia drop off in magnitude and $E(B-V)$ at peak. These parameters are sampled from the asymmetric-Gaussian distributions determined in \citet{scolnic_measuring_2016}. The distribution for $x1$ has a mean of 0.973, with $\sigma_{-} = 1.472$ and $\sigma_{+} = 0.222$. The $c$ distribution has $\mu = -0.054$, $\sigma_{-} = 0.043$, and $\sigma_{+} = 0.101$. This provides a good sampling of light curves typically observed for Type Ia SNe. The total number of CC SNe, $N_{CC}$, were divided into subtypes following \citet{eldridge_death_2013}. The only changes here are the grouping of IIb and IIpec with the IIL. See Table \ref{tab:CCsubtypesEldridge} for these rates and sub-type descriptions. 

The key differences between the analysis for expected detections done here and that carried out in S18 are the updated rates, and true survey data for cadence and depth. One other difference is in the lensing magnifications applied to the SNe; S18 used a common value of $\mu = 5$ for every one of their systems. As mentioned in their analysis this is a simple \footnote{Source-plane position being key in the true magnification of a point source object} and conservative value for the magnifications. In the analysis done here however, the target-specific total lensing magnifications determined in \citet{shu_sloan_2017} are used instead. The range covered, $\mu = [2,105] $, is consistent with their fiducial value and is in agreement with previously discovered glSNe. This is a broad range of potential magnifications, but especially with the effects of gravitational microlensing included, this range of magnifications is easily possible for all of our sources. The larger end of this distribution is also attainable if the SN happens to be close to one of the caustics from the lensing distribution, causing a large base magnification in addition to any microlensing effects.

Ideally, our simulations would marginalize over the possible magnifications, which vary as a function of the position of the SN in the source galaxy and based on the microlensing configurations. In practice however, we approximate all magnifications as the magnification estimated in S18, to reduce computational cost. The model magnifications have uncertainties of $\sim16\%$, which is insignificant compared to the uncertainties on the SN rates. Variations as a function of the position of the SN will be larger than this, and may contribute an additional source of uncertainty to our yield estimates here. Without simulating microlensing effects in combination with the lens models, it is difficult to accurately predict the scale of this uncertainty on our detection rates. Based on \cite{Dobler2006}, many supernovae can experience variations in their light curves from microlensing alone on the order of 0.5 magnitudes, but the amount of variation will depend on the lensing system.

\begin{table*}
\caption{Details of the luminosity functions and templates used for the supernova population model. Originally given in Vega magnitudes in \citet{goldstein_rates_2019} these have been converted to the AB system following the B band shift, $m_{\rm{AB}} - m_{\rm{Vega}} = 0.09$, determined in \citet{blanton_k-corrections_2007}. This table shows the mean and standard deviation for the magnitude distribution for each SN type, as well as references for additional information.}
\begin{tabular*}{\textwidth}{@{\extracolsep{\stretch{1}}}*{6}{r}}
SN Type & $\mu_{M_B}$ & $\sigma_{M_B}$ & Template & Template Reference & Luminosity Reference \\ \cmidrule(r){1-6}
Ia & -19.32 & 0.10 & Salt 2 & \citet{guy_salt2_2007} & \citet{sullivan_rates_2006}\\
IIp & -16.99 & 1.12 & SN 2005lc & \citet{sako_photometric_2011}  & \citet{li_nearby_2011}  \\
IIL & -17.55 & 0.38 & Nugent-IIL & \citet{gilliland_high_1999}  & \citet{li_nearby_2011}  \\
IIn & -19.14 & 0.50 & Nugent-IIn & \citet{gilliland_high_1999}  & \citet{li_nearby_2011} \\
Ibc & -17.60 & 0.74 & Nugent-Ibc & \citet{levan_grb_2005} & \citet{li_nearby_2011} \\ \cmidrule(r){1-6}
\end{tabular*}
\label{tab:LuminosityandTemplates}
\end{table*}

\begin{table*}
\caption{Relative rates of CC SN subtypes from \citet{eldridge_death_2013}. The IIb and IIpec fractions are grouped with IIL template in our analysis, where the Type IIL rate is the sum of the IIL, IIb and IIpec SN rates, giving a relative rate of $16.1\%$.}
\begin{tabular*}{\textwidth}{@{\extracolsep{\stretch{1}}}*{3}{r}}
CC Type & Rel. Rate ($\%$) & Descr. \\ \cmidrule(r){1-3}
IIp & 55.5 & ``plateau" of lc, H-shell interactions RSG progenitor \\
IIL & 3.0 & ``linear" decay of lc \\ 
IIn & 2.4 & ``narrow" H-lines, interactions with circumstellar material \\
IIb & 12.1 & H-lines only in early weeks, transitional to Ibc \\ 
IIpec & 1.0 & ``peculiar" 1987a like, BSG progenitor \\ 
Ib & 9.0 & No H lines, WR progenitor \\ 
Ic & 17.0 & No H, He lines, or WR progenitor \\ \cmidrule(r){1-3}
\end{tabular*}
\label{tab:CCsubtypesEldridge}
\end{table*}

\begin{figure}
    \centering
    \includegraphics[width=0.95\columnwidth]{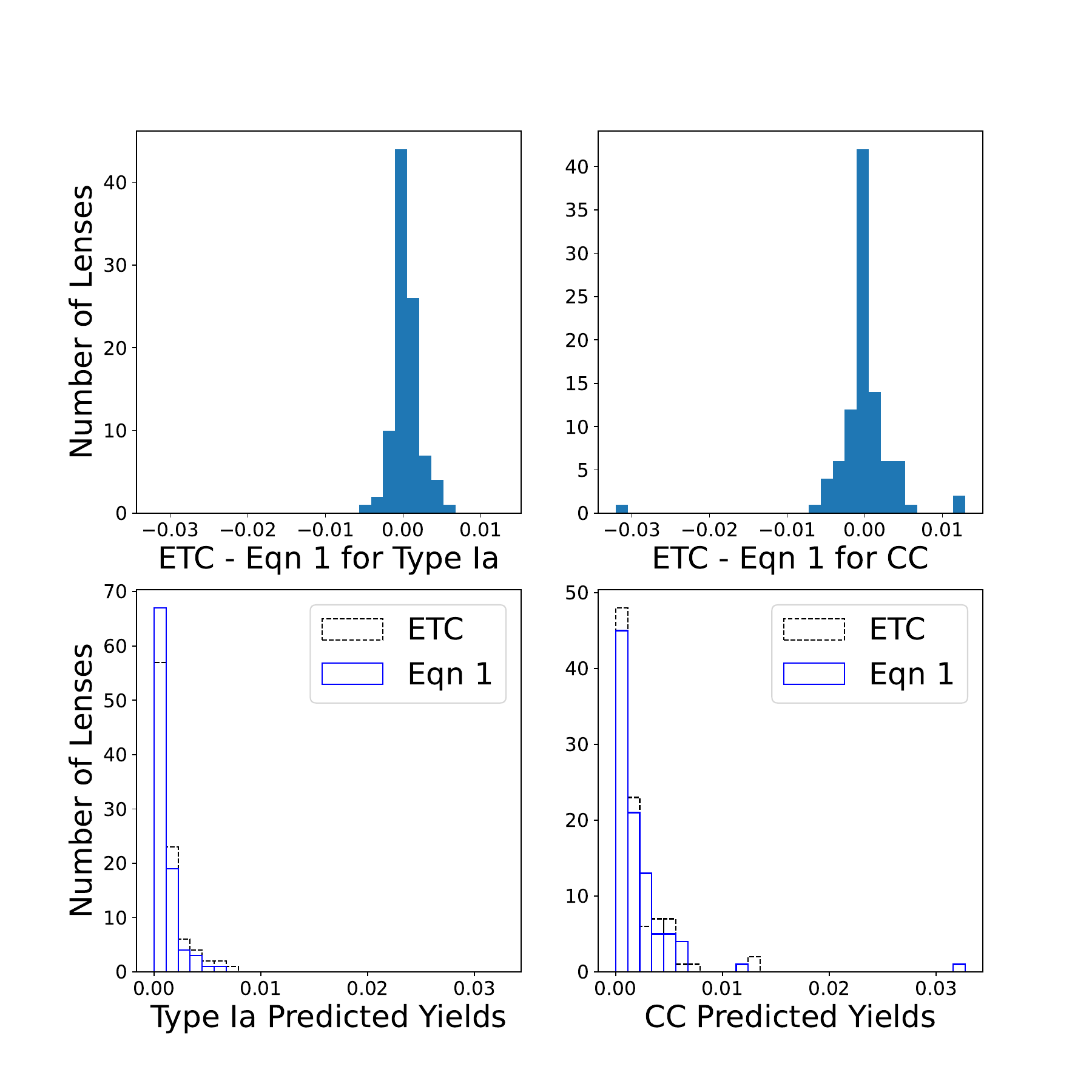}
    \caption{On top, histograms show the distributions of the difference between the ETC yield estimates and those generated with equation \ref{eq:eff}. On the bottom row, the histograms show the predicted yields of our survey per galaxy based on both \ref{eq:eff} and the ETC.}
    \label{fig:my_label}
\end{figure}

Our predicted yield is then determined by using the ETC-predicted limiting magnitudes and our internally-derived efficiency curves to count the fraction of SN explosions that would be detected by the LCOLSS. This fraction then provides our total expected SN yield. Our predicted yield over the full survey for the sources in our sample can be seen in Figure \ref{fig:my_label}. This figure also displays the differences between our derived detection efficiency and that determined by the ETC predictions.

\begin{figure*}
    \centering
    \includegraphics[width=0.95\textwidth]{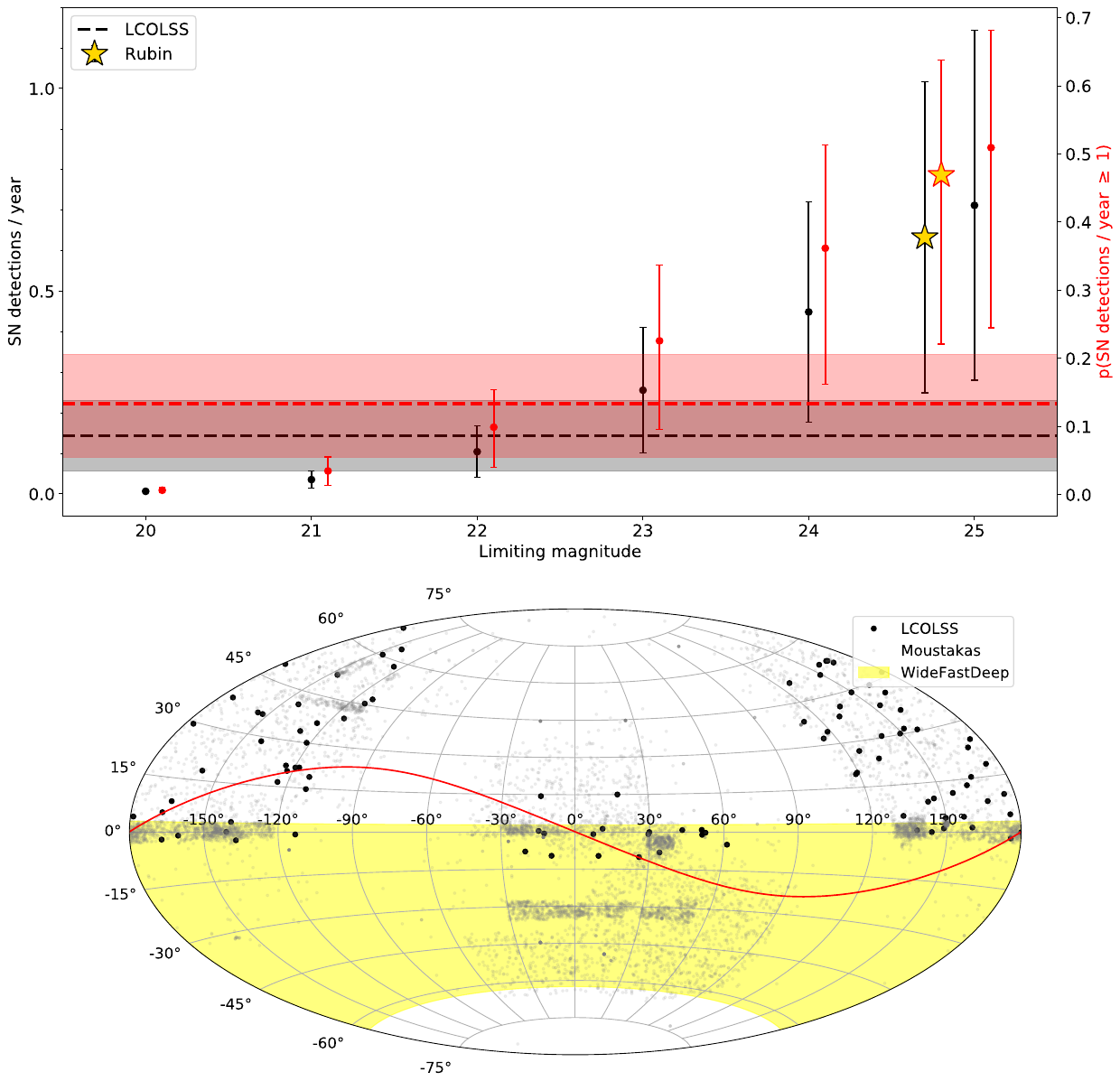}
    \caption{The top figure shows the expected yearly SN detections to come from the LCOLSS survey (in black) and the corresponding Poisson probability of this resulting in at least one SN detection per year (in red). The dashed horizontal lines show the expected value for SN detections/year, ($0.14 \pm 0.08$), using the Exposure Time Calculator. The efficiency curve method (omitted) is in very good agreement. The shaded regions in red and black, correspond to the uncertainties on the expected detections, and the Poisson probabilities at those error bounds. The circular markers show the results assuming constant single visit limiting magnitudes. The gold star provides an estimate corresponding to the values if this survey were carried out using WFD imaging planned with Rubin; single visit (2x15 second) 5 sigma r band depth of 24.7. This estimate assumes that Rubin uses our cadence and set of targets (or a similar set of targets that are within Rubin's coverage). Below is a map highlighting the region which will be surveyed with Rubin's Wide Fast Deep survey. The location of the LCOLSS targets are shown in black and a larger catalog of all known lensing systems \citep{masterlens_ref} in grey. Rubin's increased depth and exposure, along with known (and to-be discovered) lensing systems, should provide many detections over the 10-year program.}
    \label{fig:skymap_wfd}
\end{figure*}

These simulations assume single images for each SN, and do not consider any blending between adjacent images. Typical FWHM values for our survey are greater than 3 arcseonds (average seeing for our source list \ref{tab:obs}, which is more than twice the typical Einstein radius. As a result, in systems where the time delays and magnifications would allow multiple images to appear in our data simultaneously, there is a reasonable chance that multiple images would be blended. These blending effects are not expected to make a substantial impact on our yields, however, and any adjustments would lead to larger detection rates. Blended sources would have a bit more flux than the non-blended sources modelled here, and our methods would assign them a higher chance of being detected. In practice the scale of these effects is likely small, and depends on the magnifications of the different images. In general the highest magnification image will be the relevant one, as it contributes most of the flux, where the impact of blending in a fainter image will not significantly improve our detection chances unless the magnifications happen to be similar.

\section{Results} \label{sec:results}

The assumed SN data, i.e., the luminosity functions, SN rates, and relative CC rates, all have associated uncertainties. These uncertainties are accounted for in the Monte Carlo simulation we carry out to generate SN light curves. The largest uncertainties are for the supernova rates, with an average fractional uncertainty of $\frac{e_{RSN}}{R_{SN}} = 0.57 $. The SN rate uncertainty is not directly built into the Monte Carlo sims, as we directly scale up the existing SN rate measurements to determine the number of light curves to generate. However, this uncertainty, along with an additional uncertainty from the magnifications and the redshifts, is applied to our final yield estimates. This is a straightforward calculation, since the detection rate scales linearly with the SN rate.

The luminosity function uncertainty is smaller for the Type Ia events than it is for CC SNe, with errors on the peak absolute magnitudes of 0.1 and $\sim0.5$, respectively. The CC luminosity function uncertainty varies somewhat with the SN subtype \citep{Goldstein2019}. These variations are accounted for in our Monte Carlo simulations, where for a given SN event the luminosity function is randomly drawn based on these uncertainties. As a result, the simulated CC SNe have larger spreads in their peak luminosities than the Type Ia SNe do. With an uncertainty of 0.1 mag, the SN Ia distributions don't make a difference unless the observation is already very close to our detection threshold, so the effect is small. CC SNe will have greater dependence, but still need to be within 0.5 mag of the detection limit for this to be a significant effect.

Another potential uncertainty term is the uncertainty on our detection efficiency curves. The expected uncertainty on the detection efficiency, used for the purposes of fitting our detection efficiency model, is $\sim 0.1$. A change in the detection efficiency on this scale will be less relevant than the $\sim 60 \%$ uncertainty on the SN rates. For most sources, the relevant uncertainties are likely much lower than this, and only within $\sim$0.5-1 magnitudes of the central magnitude of a given detection efficiency curve do we expect there to be significant uncertainty on the detection efficiency. Sources much brighter or fainter than the central magnitude will have uncertainties close to 0. A more significant additional source of uncertainty likely stems from the potential variation in magnifications of the lensed SN, which can cover a large range of possible values. This is also most relevant for sources that are close to our detection threshold, and will vary from one system to another.

The redshifts and magnifications of the target data from S18 have smaller uncertainties. The redshifts are all measured spectroscopically, for both the lensing and host galaxies. The average lensing magnifications, $\mu$, are the ratio of total flux mapped onto the image plane to the flux in the source plane. These were determined in \citet{shu_sloan_2017} using singular isothermal ellipsoid (SIE) modeling with the HST F814W images of the lensed galaxies. The magnifications determined using the modeling done in \citet{shu_sloan_2017}, were compared to the values from previous models done for the 25 BELLS strong lens systems in \citet{brownstein_boss_2011},  $\mu_{\rm{ratio}} = \mu_{\rm{shu2018}}/\mu_{\rm{brownstein2012}}$. The average ratio, $\bar{\mu_{\rm}} = 0.94$, showed good agreement between the independent modeling. The standard deviation of the ratios, $\sigma \mu_{\rm{ratio}} = 0.16$ is used as the fractional uncertainty for all the lens modeling. The uncertainties here are $\sim$4 times smaller than those on the SNRates. Combined uncertainties between the lens models and the supernova rates remain under $60\%$, which is the assumed uncertainty on our final yield estimates.

\subsection{Expected glSN yield for LCOLSS}

\begin{table*}
\caption{Total SN detections expected over our 2-year survey with LCO in rp (and yearly); Along with associated probabilities for at least one detection. The numbers assuming various limiting magnitude depths as constants for each observation included as well.}
\begin{tabular*}{\textwidth}{@{\extracolsep{\stretch{1}}}*{3}{r}}
Method & Detections: Total (Yearly) $\overline{SN}$,$\overline{Ia}$,$\overline{CC}$ & p(SN Detections $\geq$ 1): Total (Yearly) \\ \cmidrule(r){1-3}
Exposure Time Calculator & 0.198,0.080,0.117 (0.092,0.037,0.054) & 0.179 (0.088) \\
Efficiency Curve & 0.192,0.062,0.13 (0.089,0.029,0.060) & 0.174 (0.085) \\ 
$20$-limiting & 0.009,0.005,0.004 (0.004,0.002,0.002) & 0.01 (0.004) \\
$21$-limiting & 0.049,0.026,0.023 (0.023,0.012,0.011) & 0.048 (0.023) \\
$22$-limiting & 0.141,0.065,0.076 (0.065,0.030,0.035) & 0.131 (0.063) \\
$23$-limiting & 0.357,0.142,0.215 (0.165,0.066,0.010) & 0.300 (0.152)  \\
$24$-limiting & 0.631,0.173,0.458 (0.292,0.080,0.212) & 0.468 (0.253) \\
$25$-limiting & 0.987,0.259,0.728 (0.457,0.120,0.337) & 0.627 (0.367) \\\cmidrule(r){1-3}
\end{tabular*}
\label{tab:Detections}
\end{table*}

For the LCOLSS sample, we find an expected yield and $68\%$ Poisson confidence intervals, $N_{SN} = 0.20, [0,2.1] $, $N_{Ia} = 0.08, [0,2.0]$, $N_{CC} = 0.12, [0,2.0]$ of total SNe, Type Ia, and core-collapse SNe respectively. Based on these yield estimates, the null result of this survey is not surprising. After correcting the predicted SFRs in the target lensed galaxies, and accounting for the depth achievable with 1-meter class telescopes, we find that the probability of detecting at least one 1 lensed SN per year with this survey design is less than 0.20. The key improvement for future targeted SN surveys to be successful would be to reach fainter magnitudes with each observation. The predicted detection rates for our survey with a variety of different limiting magnitudes can be seen in Table \ref{tab:Detections}. For example, the 8.4-meter Rubin Observatory will reach an r-band $5\sigma$ depth of 24.7 AB mag. Observing a similar sample of $\sim 100$ strong lensing systems, Rubin Observatory will have an approximately $50\%$ chance of detecting a lensed SN each year from that sample. Access to the i and z band filters with Rubin may also help to improve the detection rate of glSNe \citep{Goldstein2019}.

Our analysis can also be used to investigate the trade-off between the limiting magnitude and the cadence. We find that the image depth plays a larger role than the cadence on the total yield. The results indicate that changing the cadence from two weeks to four weeks produces the same yield if we increase the image depth by about 0.4 magnitudes. Thus, a deeper survey can provide a better yield even with less frequent observations. However, such a survey is more likely to miss the peak of the SN, and will often have fewer observations by the time follow-up resources get triggered. This can be a large downside, as the quality of the resulting H$_0$ measurements improve significantly if the SN is detected before it peaks. Future surveys, such as the Vera C. Rubin Observatory's Legacy Survey of Space and Time (LSST), are anticipated to have substantially deeper limiting magnitudes compared to our survey, potentially indicating a significantly larger detection rate over a similar sample of lenses.

Figure \ref{fig:skymap_wfd} shows the expected yearly detection rate in our survey for a variety of limiting magnitudes. This figure also includes the values expected if the Rubin observatory were used instead of LCO for this survey, reaching deeper limiting magnitudes. In practice, Rubin observatory will not cover much of our source list due to the locations on the sky of our systems. The sky coverage of Rubin compared to our survey and the masterlens catalog can be seen in the bottom panel of Figure \ref{fig:skymap_wfd}. In addition to the number of detections per year, this figure includes the Poisson probability of getting a detection in one year of the LCOLSS survey for each magnitude. The uncertainties shown are estimated to be $60\%$ of the detection rate, which is dominated by the uncertainty on the supernova rates. This is determined as the average rate on the SN rates, combined with uncertainties on the magnifications and spectroscopic redshifts. The uncertainties on other quantities are not included here, as they are small enough compared to the uncertainties on the underlying SN rates that they only cause minimal increases to the final uncertainties.

\begin{figure}
    \centering
    \includegraphics[width=0.95\columnwidth]{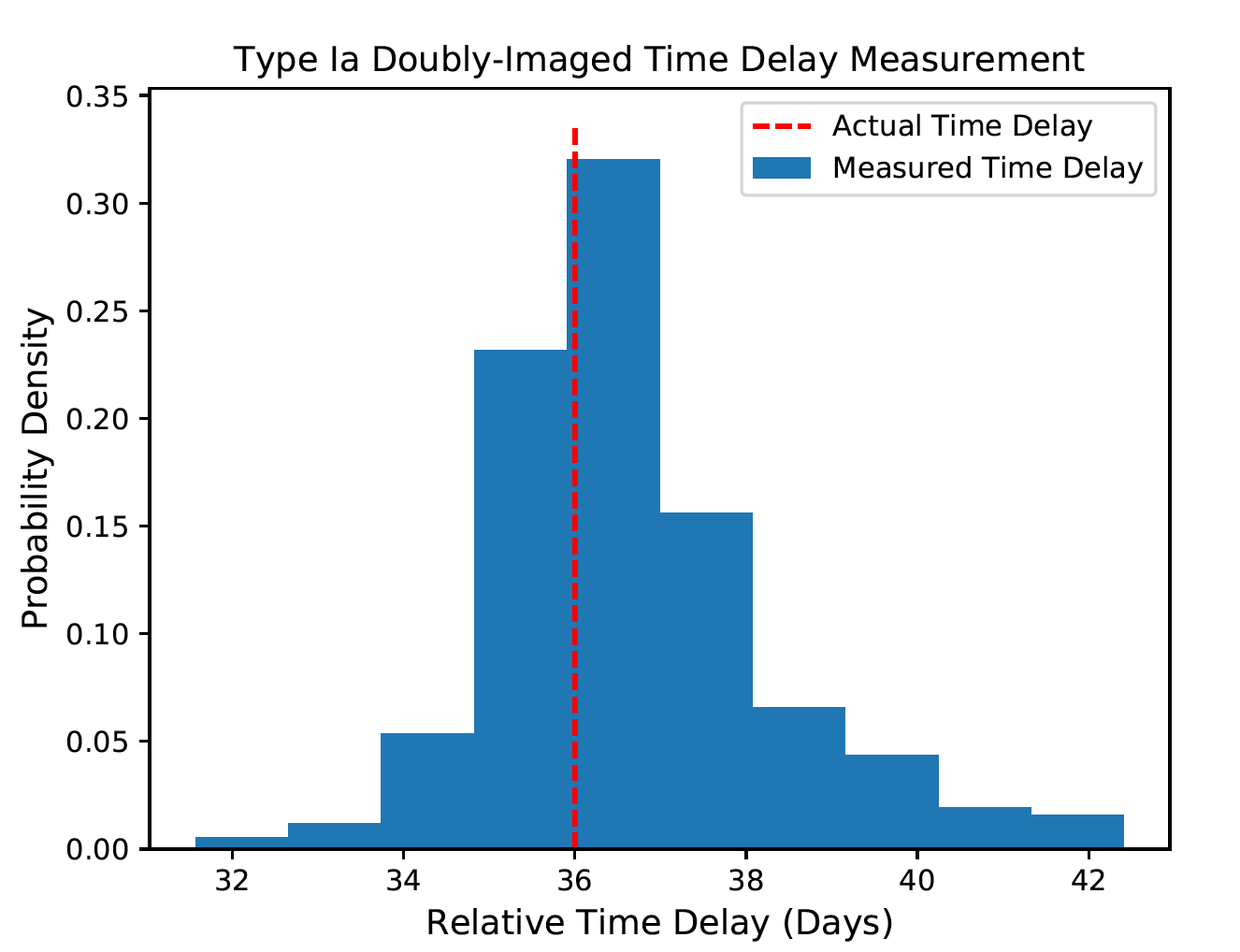}
    \caption{Histogram of expected time delays for the lensing systems in our sample. The average time delay for this sample is about 36 days.}
    \label{fig:time_delay}
\end{figure}

We have also analyzed the expected time delays for the LCOLSS sample. The average expected time delays for this sample is about 36 days, with a distribution that can be seen in Figure \ref{fig:time_delay}. The precision on the time delays with this data alone is about 5\% and 7\% for Type Ia and core-collapse SNe respectively (and therefore a roughly comparable uncertainty on $H_{0}$ assuming the uncertainty on the lens models does not exceed the precision on the time delays) if a detection is guaranteed before the peak. Here, we have followed the formalism in \cite{Pierel_Rodney} to calculate time-delays for our sample. The current and future high cadence transient surveys, the Zwicky Transient Facility (ZTF) and the Vera C. Rubin Observatory Legacy Survey of Space and Time (LSST), are/will be greatly increasing SN detection numbers. In addition to these, next generation space telescopes, particularly the Roman Space Telescope, are expected to discover large numbers of SNe at higher redshifts \citep{Pierel2021}. 

The detection efficiencies were thoroughly investigated through two different methods in the LCOLSS analysis done here; first using the Exposure Time Calculator and again with Equation \ref{eq:eff} calibrated on SDSS data. The Exposure Time Calculator counted detections in binary, based on the S/N=3 threshold predicted according to image lunar conditions and airmass. In Equation \ref{eq:eff} the detections were counted as a fractional chance of recovery, measured on the difference images. The similar yields predicted through both channels suggest that the depths of the images are well accounted for. Our ability to detect transients has been further validated by the detections of many solar system bodies during our search.
\section{Summary and Discussion} \label{sec:summary}

We find an expected number of detections and $68\%$ Poisson confidence intervals, $N_{SN} = 0.20, [0,2.1] $, $N_{Ia} = 0.08, [0,2.0]$, $N_{CC} = 0.12, [0,2.0]$, for all SN, Type Ia, and CC SNe respectively for LCOLSS per year. Thus, our analysis indicates that the null result of the LCOLSS survey is not surprising, and our updated yields are significantly lower than our pre-survey estimates. We estimate that the detection rate based on our data is relatively low, such that a significantly longer survey time or deeper limiting magnitudes for the survey would be required in order to expect a detection with this configuration. We also estimated the expectation for the SN yield if the cadence were to be lowered to once per month, instead of our standard two-week cadence. Such a strategy would produce the same lensed SN yield if the limiting magnitude were increased by 0.4 magnitudes, indicating that the depth of the survey is the dominant factor in determining the lensed SN yield. For the same survey time (and cadence), detecting fainter SNe would make the largest difference to the SN yields for a targeted survey. For a targeted search such as this one, follow-up of possible SN candidates can be more straightforwardly allocated resources on the basis of the quality of the difference images, given that we know that the source is lensed. This more straightforward correspondence between expected SN yields and actual detections suggests that discovering a lensed SN may be more easily realized in a targeted search. 

Although the LCOLSS did not produce any SN detections, our analysis here shows that a targeted survey at fainter limiting magnitudes is likely to yield a detection over one year timescales. This dominant dependence on the limiting magnitude suggests that detections of lensed SNe are more likely achieved on larger telescopes even if the observations are more infrequent ($\sim$ once a month) than our adopted cadence. LSST is forecast to be an ideal survey for searching for glSNe, and the deeper magnitude limits for LSST are quite promising for a significant detection rate of glSNe.

\bigskip
\bigskip 
\section*{Acknowledgements}

SC gratefully acknowledges support from the RCSA Time Domain Astrophysics Scialog award, NASA ATP NNX17AK90G, NSF AAG 2009574, and the IBM Einstein Fellowship from the Institute for Advanced Study. SC also gratefully acknowledges time awards LCO2020B-015, LCO2020A-019, LCO2019B-022, LCO2019A-008 on which this work is based. IPF acknowledges support from the Spanish State Research Agency
(AEI) under grant numbers ESP2017-86852-C4-2-R and PID2019-
105552RB- C43. We thank Yiping Shu, Leonidas Moustakas and Peter Nugent for many helpful discussions.

\section*{Data Availability}
The data underlying this article will be shared on reasonable request to the corresponding author.
\bigskip

\bibliographystyle{mnras}
\bibliography{SN_refs}

\appendix
\section{Details for our Target Systems}\label{app:lens_data}
\onecolumn

\setcounter{table}{0}
\renewcommand{\thetable}{A.\arabic{table}}

\setcounter{figure}{0}
\renewcommand{\thefigure}{A.\arabic{figure}}

\begin{figure}
    \centering
    
    \includegraphics[width = 0.95\columnwidth]{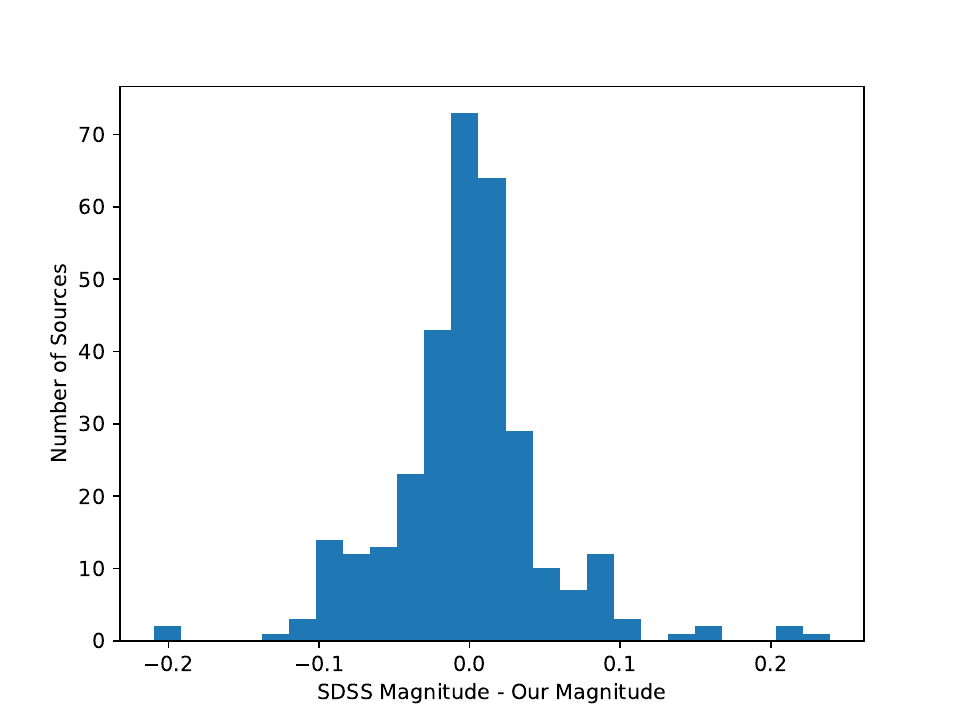}
    \caption{Sample zero-point calibration data for a single data frame. The histogram shows the distribution of the differences between our measured magnitudes and those displayed in the SDSS catalog. This indicates that the image zero-point has been correctly determined and that our measurements for the photometry are reasonably accurate.}\label{phot_zp}
\end{figure}

\begin{figure*}
    \centering
    \includegraphics[width=0.95\textwidth]{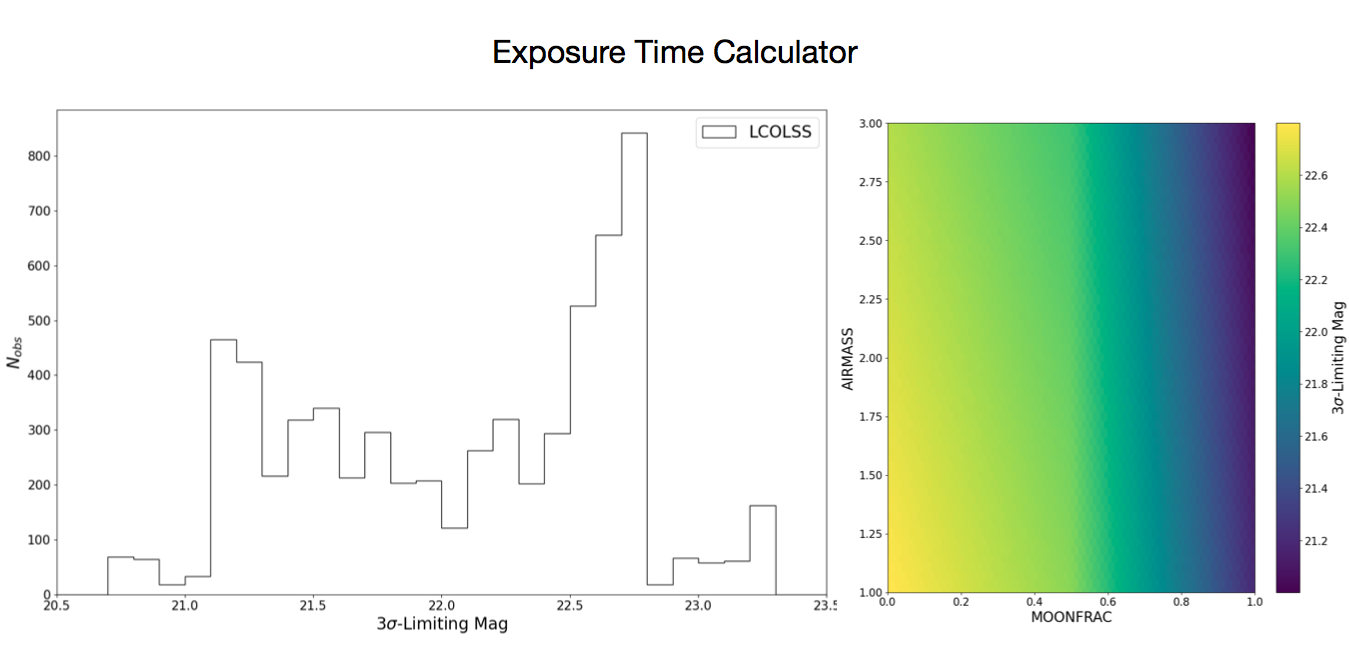}
    \caption{Limiting magnitudes predicted by the LCO Exposure Time Calculator. The figure on the left shows the distribution of $3\sigma$ limiting magnitudes for the observations in our survey. On the right the color bar shows the full range of predicted $3\sigma$ limiting magnitudes. The predictions are made using an interpolation over the lunar phase and airmass.}
    \label{fig:etc}
\end{figure*}

\setcounter{table}{0}
\renewcommand{\thetable}{A\arabic{table}}
\begin{longtable}{p{.175\textwidth}cccccccccc}
\caption{The 98 target systems included in \protect\cite{shu_prediction_2018}, with spectroscopic redshifts for lensing (z$_L$) and source (z$_S$) galaxies, magnifications ($\mu$), star formation rates [msol/yr] (SFR), and (observer-frame) supernova rates (both N$_{cc}$, the core-collapse SN rate and N$_{Ia}$, the Type Ia SN rate) [year$^{-1}$], with uncertainties. We also include the Einstein radii (R$_e$) for the available models of these systems [arcseconds]. All of these values, except for the Einstein radii, are from \protect\citep{shu_prediction_2018}, with the correction factor of $(1+z_s)^{-2}$ applied to his star formation rates. The Einstein radii are from \protect\cite{Bolton2008}, \protect\cite{Brownstein2012}, and \protect\cite{Shu2017}.}\label{tab:sources}
\xspace\\
\hline
\textbf{Survey Name}  & \textbf{z$_L$} & \textbf{z$_S$} & \textbf{$\mu$} & \textbf{SFR} & \textbf{e$_{SFR}$} & \textbf{N$_{cc}$} & \textbf{e$_{Ncc}$} & \textbf{N$_{Ia}$} & \textbf{e$_{NIa}$} & \textbf{R$_e$} \\
\hline
   SDSSJ0029-0055 &  0.23 &  0.93 &  23 &       1.42 &          0.81 &    5.1e-03 &       2.9e-03 &    6.4e-04 &       3.6e-04 &  2.16  \\ 
   SDSSJ0037-0942 &  0.20 &  0.63 &   6 &       2.37 &          1.20 &    1.0e-02 &       5.2e-03 &    1.5e-03 &       7.5e-04 &  2.19  \\ 
   SDSSJ0044+0113 &  0.12 &  0.20 &   2 &       1.74 &          0.83 &    1.0e-02 &       4.8e-03 &    1.9e-03 &       9.1e-04 &  2.61  \\ 
   SDSSJ0109+1500 &  0.29 &  0.52 &   2 &       2.64 &          1.64 &    1.2e-02 &       7.6e-03 &    1.9e-03 &       1.2e-03 &  1.38  \\ 
   SDSSJ0157-0056 &  0.51 &  0.92 &   2 &      22.76 &         16.33 &    8.3e-02 &       5.9e-02 &    1.0e-02 &       7.4e-03 &  1.06  \\ 
   SDSSJ0216-0813 &  0.33 &  0.52 &   3 &       8.48 &          4.37 &    3.9e-02 &       2.0e-02 &    6.0e-03 &       3.1e-03 &  2.67  \\ 
   SDSSJ0252+0039 &  0.28 &  0.98 &  16 &       4.90 &          2.24 &    1.7e-02 &       7.9e-03 &    2.1e-03 &       9.6e-04 &  1.39  \\ 
   SDSSJ0330-0020 &  0.35 &  1.07 &   4 &      22.94 &         11.16 &    7.7e-02 &       3.8e-02 &    9.0e-03 &       4.4e-03 &  1.2  \\ 
   SDSSJ0405-0455 &  0.08 &  0.81 &  16 &       0.64 &          0.31 &    2.5e-03 &       1.2e-03 &    3.3e-04 &       1.6e-04 &  1.36  \\ 
   SDSSJ0728+3835 &  0.21 &  0.69 &  10 &       0.98 &          0.67 &    4.0e-03 &       2.7e-03 &    5.7e-04 &       3.9e-04 &  1.78  \\ 
   SDSSJ0737+3216 &  0.32 &  0.58 &  14 &       1.84 &          0.92 &    8.1e-03 &       4.1e-03 &    1.2e-03 &       6.1e-04 &  2.82  \\ 
   SDSSJ0822+2652 &  0.24 &  0.59 &   7 &       1.27 &          0.63 &    5.6e-03 &       2.8e-03 &    8.2e-04 &       4.1e-04 &  1.82  \\ 
 SWELLSJ0841+3824 &  0.12 &  0.66 &   5 &     -99.0 &        -99.0 &   -99.0 &      -99.0 &   -99.0 &      -99.0 &  4.21  \\ 
   SDSSJ0903+4116 &  0.43 &  1.06 &   8 &       6.34 &          4.88 &    2.1e-02 &       1.7e-02 &    2.5e-03 &       1.9e-03 &  1.78  \\ 
   SDSSJ0912+0029 &  0.16 &  0.32 &   6 &       1.43 &          0.92 &    7.6e-03 &       4.9e-03 &    1.3e-03 &       8.4e-04 &  3.87  \\ 
   SDSSJ0935-0003 &  0.35 &  0.47 &   3 &       3.19 &          3.05 &    1.5e-02 &       1.4e-02 &    2.4e-03 &       2.3e-03 &  4.24  \\ 
   SDSSJ0936+0913 &  0.19 &  0.59 &   7 &       1.19 &          0.63 &    5.2e-03 &       2.8e-03 &    7.7e-04 &       4.1e-04 &  2.11  \\ 
   SDSSJ0946+1006 &  0.22 &  0.61 &  18 &       1.77 &          0.93 &    7.7e-03 &       4.0e-03 &    1.1e-03 &       5.9e-04 &  2.35  \\ 
   SDSSJ0955+0101 &  0.11 &  0.32 &   4 &       1.21 &          0.57 &    6.4e-03 &       3.0e-03 &    1.1e-03 &       5.3e-04 &  1.09  \\ 
   SDSSJ0956+5100 &  0.24 &  0.47 &   8 &       1.34 &          0.74 &    6.4e-03 &       3.5e-03 &    1.0e-03 &       5.6e-04 &  2.19  \\ 
   SDSSJ0959+0410 &  0.13 &  0.54 &   5 &       2.07 &          1.31 &    9.4e-03 &       5.9e-03 &    1.4e-03 &       9.0e-04 &  1.39  \\ 
   SDSSJ0959+4416 &  0.24 &  0.53 &   9 &       0.81 &          0.43 &    3.7e-03 &       1.9e-03 &    5.7e-04 &       3.0e-04 &  1.98  \\ 
   SDSSJ1016+3859 &  0.17 &  0.44 &   5 &       0.77 &          0.48 &    3.7e-03 &       2.3e-03 &    6.0e-04 &       3.8e-04 &  1.46  \\ 
   SDSSJ1020+1122 &  0.28 &  0.55 &   6 &       1.37 &          0.79 &    6.2e-03 &       3.6e-03 &    9.4e-04 &       5.4e-04 &  1.59  \\ 
 SWELLSJ1029+0420 &  0.10 &  0.62 &   4 &       2.90 &          1.64 &    1.2e-02 &       7.1e-03 &    1.8e-03 &       1.0e-03 &  1.56  \\ 
   SDSSJ1032+5322 &  0.13 &  0.33 &   3 &       3.39 &          1.58 &    1.8e-02 &       8.3e-03 &    3.1e-03 &       1.4e-03 &  0.81  \\ 
   SDSSJ1100+5329 &  0.32 &  0.86 &  13 &       8.87 &          4.77 &    3.3e-02 &       1.8e-02 &    4.3e-03 &       2.3e-03 &  2.24  \\ 
   SDSSJ1103+5322 &  0.16 &  0.74 &   3 &      25.20 &         18.33 &    1.0e-01 &       7.4e-02 &    1.4e-02 &       1.0e-02 &  1.95  \\ 
   SDSSJ1106+5228 &  0.10 &  0.41 &  25 &       1.16 &          0.70 &    5.7e-03 &       3.5e-03 &    9.4e-04 &       5.7e-04 &  1.68  \\ 
   SDSSJ1112+0826 &  0.27 &  0.63 &   4 &      12.04 &          6.62 &    5.2e-02 &       2.8e-02 &    7.5e-03 &       4.1e-03 &  1.5  \\ 
   SDSSJ1142+1001 &  0.22 &  0.50 &   3 &       2.27 &          1.20 &    1.1e-02 &       5.6e-03 &    1.6e-03 &       8.7e-04 &  1.91  \\ 
   SDSSJ1143-0144 &  0.11 &  0.40 &   4 &       8.47 &          3.93 &    4.2e-02 &       2.0e-02 &    7.0e-03 &       3.2e-03 &  4.8  \\ 
   SDSSJ1153+4612 &  0.18 &  0.88 &  10 &       2.21 &          1.13 &    8.2e-03 &       4.2e-03 &    1.0e-03 &       5.3e-04 &  1.16  \\ 
   SDSSJ1204+0358 &  0.16 &  0.63 &   9 &       2.60 &          1.39 &    1.1e-02 &       6.0e-03 &    1.6e-03 &       8.6e-04 &  1.47  \\ 
   SDSSJ1205+4910 &  0.21 &  0.48 &  13 &       1.23 &          0.55 &    5.8e-03 &       2.6e-03 &    9.1e-04 &       4.1e-04 &  2.59  \\ 
   SDSSJ1213+6708 &  0.12 &  0.64 &   8 &       2.75 &          1.45 &    1.2e-02 &       6.2e-03 &    1.7e-03 &       8.9e-04 &  3.23  \\ 
   SDSSJ1250+0523 &  0.23 &  0.80 &  10 &       5.43 &          2.87 &    2.1e-02 &       1.1e-02 &    2.8e-03 &       1.5e-03 &  1.81  \\ 
   SDSSJ1251-0208 &  0.22 &  0.78 &   4 &       6.44 &          3.47 &    2.5e-02 &       1.4e-02 &    3.4e-03 &       1.8e-03 &  2.61  \\ 
   SDSSJ1416+5136 &  0.30 &  0.81 &   4 &      16.48 &         10.59 &    6.4e-02 &       4.1e-02 &    8.4e-03 &       5.4e-03 &  1.43  \\ 
   SDSSJ1420+6019 &  0.06 &  0.54 &  13 &       1.64 &          0.84 &    7.5e-03 &       3.8e-03 &    1.1e-03 &       5.8e-04 &  2.06  \\ 
   SDSSJ1430+4105 &  0.28 &  0.58 &   6 &       7.41 &          3.69 &    3.3e-02 &       1.6e-02 &    4.9e-03 &       2.4e-03 &  2.55  \\ 
   SDSSJ1432+6317 &  0.12 &  0.66 &   4 &      16.44 &          7.80 &    6.9e-02 &       3.3e-02 &    9.9e-03 &       4.7e-03 &  5.85  \\ 
   SDSSJ1436-0000 &  0.29 &  0.80 &   4 &       5.83 &          3.09 &    2.3e-02 &       1.2e-02 &    3.0e-03 &       1.6e-03 &  2.24  \\ 
   SDSSJ1443+0304 &  0.13 &  0.42 &   7 &       0.45 &          0.25 &    2.2e-03 &       1.2e-03 &    3.6e-04 &       2.0e-04 &  0.94  \\ 
   SDSSJ1451-0239 &  0.13 &  0.52 &  11 &       1.43 &          0.74 &    6.6e-03 &       3.4e-03 &    1.0e-03 &       5.2e-04 &  2.48  \\ 
   SDSSJ1525+3327 &  0.36 &  0.72 &   4 &       7.98 &          3.85 &    3.2e-02 &       1.6e-02 &    4.5e-03 &       2.2e-03 &  2.9  \\ 
   SDSSJ1538+5817 &  0.14 &  0.53 &   8 &       7.05 &          3.20 &    3.2e-02 &       1.5e-02 &    4.9e-03 &       2.2e-03 &  1.58  \\ 
   SDSSJ1621+3931 &  0.24 &  0.60 &   8 &       5.27 &          4.92 &    2.3e-02 &       2.1e-02 &    3.4e-03 &       3.2e-03 &  2.14  \\ 
   SDSSJ1627-0053 &  0.21 &  0.52 &  20 &       0.56 &          0.26 &    2.6e-03 &       1.2e-03 &    4.0e-04 &       1.8e-04 &  1.98  \\ 
   SDSSJ1630+4520 &  0.25 &  0.79 &   9 &       5.21 &          3.00 &    2.0e-02 &       1.2e-02 &    2.7e-03 &       1.6e-03 &  1.96  \\ 
   SDSSJ1636+4707 &  0.23 &  0.67 &   8 &       0.82 &          0.47 &    3.4e-03 &       1.9e-03 &    4.9e-04 &       2.8e-04 &  1.68  \\ 
   SDSSJ2238-0754 &  0.14 &  0.71 &  12 &       2.53 &          1.30 &    1.0e-02 &       5.3e-03 &    1.4e-03 &       7.4e-04 &  2.33  \\ 
   SDSSJ2300+0022 &  0.23 &  0.46 &  12 &       0.33 &          0.23 &    1.6e-03 &       1.1e-03 &    2.5e-04 &       1.8e-04 &  1.83  \\ 
   SDSSJ2303+1422 &  0.16 &  0.52 &   8 &       0.69 &          0.48 &    3.2e-03 &       2.2e-03 &    4.9e-04 &       3.4e-04 &  3.28  \\ 
   SDSSJ2321-0939 &  0.08 &  0.53 &  12 &       2.86 &          1.41 &    1.3e-02 &       6.4e-03 &    2.0e-03 &       9.8e-04 &  4.11  \\ 
   SDSSJ0143-1006 &  0.22 &  1.10 &   3 &      36.17 &         22.47 &    1.2e-01 &       7.5e-02 &    1.4e-02 &       8.6e-03 &  1.23  \\ 
   SDSSJ0159-0006 &  0.16 &  0.75 &   6 &       1.70 &          0.95 &    6.8e-03 &       3.8e-03 &    9.2e-04 &       5.1e-04 &  0.92  \\ 
   SDSSJ0324+0045 &  0.32 &  0.92 &  14 &       0.79 &          0.49 &    2.9e-03 &       1.8e-03 &    3.6e-04 &       2.2e-04 &  0.55  \\ 
   SDSSJ0324-0110 &  0.45 &  0.62 &   4 &       1.33 &          0.65 &    5.7e-03 &       2.8e-03 &    8.4e-04 &       4.1e-04 &  0.63  \\ 
   SDSSJ0753+3416 &  0.14 &  0.96 &  27 &       2.92 &          1.64 &    1.0e-02 &       5.8e-03 &    1.3e-03 &       7.2e-04 &  1.23  \\ 
   SDSSJ0754+1927 &  0.15 &  0.74 &   5 &       5.91 &          3.60 &    2.4e-02 &       1.4e-02 &    3.2e-03 &       2.0e-03 &  1.04  \\ 
   SDSSJ0757+1956 &  0.12 &  0.83 &   9 &       8.48 &          4.48 &    3.2e-02 &       1.7e-02 &    4.2e-03 &       2.2e-03 &  1.62  \\ 
   SDSSJ0847+2348 &  0.16 &  0.53 &  17 &       0.34 &          0.17 &    1.6e-03 &       7.8e-04 &    2.4e-04 &       1.2e-04 &  0.96  \\ 
   SDSSJ0851+0505 &  0.13 &  0.64 &   6 &       3.09 &          1.67 &    1.3e-02 &       7.1e-03 &    1.9e-03 &       1.0e-03 &  0.91  \\ 
   SDSSJ0920+3028 &  0.29 &  0.39 &   8 &       0.57 &          0.26 &    2.9e-03 &       1.3e-03 &    4.7e-04 &       2.2e-04 &  0.7  \\ 
   SDSSJ0955+3014 &  0.32 &  0.47 &   4 &       0.83 &          0.83 &    4.0e-03 &       4.0e-03 &    6.3e-04 &       6.3e-04 &  0.54  \\ 
   SDSSJ1010+3124 &  0.17 &  0.42 &   4 &       2.43 &          1.24 &    1.2e-02 &       6.1e-03 &    1.9e-03 &       9.9e-04 &  1.14  \\ 
   SDSSJ1031+3026 &  0.17 &  0.75 &   5 &       6.53 &          3.40 &    2.6e-02 &       1.4e-02 &    3.5e-03 &       1.8e-03 &  0.88  \\ 
   SDSSJ1040+3626 &  0.12 &  0.28 &   3 &       0.49 &          0.24 &    2.7e-03 &       1.3e-03 &    4.7e-04 &       2.4e-04 &  0.59  \\ 
   SDSSJ1041+0112 &  0.10 &  0.22 &   5 &       0.20 &          0.13 &    1.2e-03 &       7.7e-04 &    2.1e-04 &       1.4e-04 &  0.6  \\ 
   SDSSJ1048+1313 &  0.13 &  0.67 &   4 &       6.56 &          6.53 &    2.7e-02 &       2.7e-02 &    3.9e-03 &       3.9e-03 &  1.18  \\ 
   SDSSJ1051+4439 &  0.16 &  0.54 &   3 &       7.29 &          3.71 &    3.3e-02 &       1.7e-02 &    5.0e-03 &       2.6e-03 &  0.99  \\ 
   SDSSJ1056+4141 &  0.13 &  0.83 &  10 &       1.85 &          0.87 &    7.1e-03 &       3.3e-03 &    9.2e-04 &       4.3e-04 &  0.72  \\ 
   SDSSJ1101+1523 &  0.18 &  0.52 &   5 &       1.51 &          0.82 &    7.0e-03 &       3.8e-03 &    1.1e-03 &       5.8e-04 &  1.18  \\ 
   SDSSJ1127+2312 &  0.13 &  0.36 &   8 &       2.27 &          1.08 &    1.2e-02 &       5.5e-03 &    2.0e-03 &       9.4e-04 &  1.25  \\ 
   SDSSJ1137+1818 &  0.12 &  0.46 &  10 &       1.74 &          0.84 &    8.3e-03 &       4.0e-03 &    1.3e-03 &       6.4e-04 &  1.29  \\ 
   SDSSJ1142+2509 &  0.16 &  0.66 &  18 &       0.58 &          0.29 &    2.4e-03 &       1.2e-03 &    3.5e-04 &       1.7e-04 &  0.79  \\ 
   SDSSJ1144+0436 &  0.10 &  0.26 &   5 &       0.94 &          0.50 &    5.2e-03 &       2.8e-03 &    9.4e-04 &       5.0e-04 &  0.76  \\ 
   SDSSJ1213+2930 &  0.09 &  0.60 &  21 &       0.78 &          0.43 &    3.4e-03 &       1.9e-03 &    5.0e-04 &       2.8e-04 &  1.35  \\ 
   SDSSJ1301+0834 &  0.09 &  0.53 &   9 &       2.48 &          2.14 &    1.1e-02 &       9.7e-03 &    1.7e-03 &       1.5e-03 &  1.0  \\ 
   SDSSJ1330+1750 &  0.21 &  0.37 &   4 &       1.01 &          0.48 &    5.2e-03 &       2.4e-03 &    8.7e-04 &       4.1e-04 &  1.01  \\ 
   SDSSJ1433+2835 &  0.09 &  0.41 &  10 &       1.11 &          0.55 &    5.5e-03 &       2.7e-03 &    9.0e-04 &       4.5e-04 &  1.53  \\ 
   SDSSJ1541+3642 &  0.14 &  0.74 &  16 &       2.28 &          1.06 &    9.1e-03 &       4.2e-03 &    1.2e-03 &       5.8e-04 &  1.17  \\ 
   SDSSJ1543+2202 &  0.27 &  0.40 &   3 &       3.42 &          1.58 &    1.7e-02 &       7.9e-03 &    2.8e-03 &       1.3e-03 &  0.78  \\ 
   SDSSJ1550+2020 &  0.14 &  0.35 &   2 &       3.95 &          1.98 &    2.0e-02 &       1.0e-02 &    3.5e-03 &       1.7e-03 &  1.01  \\ 
   SDSSJ1553+3004 &  0.16 &  0.57 &   5 &       2.23 &          1.10 &    9.9e-03 &       4.9e-03 &    1.5e-03 &       7.3e-04 &  0.84  \\ 
   SDSSJ1607+2147 &  0.21 &  0.49 &   2 &       2.07 &          1.17 &    9.7e-03 &       5.5e-03 &    1.5e-03 &       8.6e-04 &  0.57  \\ 
   SDSSJ1633+1441 &  0.13 &  0.58 &  19 &       1.04 &          0.52 &    4.6e-03 &       2.3e-03 &    6.8e-04 &       3.4e-04 &  1.39  \\ 
   SDSSJ2309-0039 &  0.29 &  1.00 &   4 &      11.53 &          8.07 &    4.0e-02 &       2.8e-02 &    4.8e-03 &       3.4e-03 &  1.14  \\ 
  BELLSJ0830+5116 &  0.53 &  1.33 &   6 &      24.68 &         15.93 &    7.4e-02 &       4.8e-02 &    7.6e-03 &       4.9e-03 &  1.142  \\ 
  BELLSJ1159-0007 &  0.58 &  1.35 &   8 &      14.74 &          7.03 &    4.4e-02 &       2.1e-02 &    4.5e-03 &       2.1e-03 &  0.683  \\ 
  BELLSJ1221+3806 &  0.53 &  1.28 &   4 &      18.70 &         10.85 &    5.7e-02 &       3.3e-02 &    6.0e-03 &       3.5e-03 &  0.699  \\ 
  BELLSJ1318-0104 &  0.66 &  1.40 &   7 &      11.86 &          6.30 &    3.4e-02 &       1.8e-02 &    3.5e-03 &       1.8e-03 &  0.679  \\ 
  BELLSJ1337+3620 &  0.56 &  1.18 &  12 &       9.76 &          6.12 &    3.1e-02 &       2.0e-02 &    3.5e-03 &       2.2e-03 &  1.386  \\ 
  BELLSJ1349+3612 &  0.44 &  0.89 &   5 &       6.61 &          3.14 &    2.4e-02 &       1.2e-02 &    3.1e-03 &       1.5e-03 &  0.75  \\ 
  BELLSJ1542+1629 &  0.35 &  1.02 &   3 &      14.02 &         11.27 &    4.8e-02 &       3.9e-02 &    5.8e-03 &       4.6e-03 &  1.042  \\ 
  BELLSJ1601+2138 &  0.54 &  1.45 &   3 &      25.22 &         19.89 &    7.2e-02 &       5.7e-02 &    7.1e-03 &       5.6e-03 &  0.858  \\ 
  BELLSJ1631+1854 &  0.41 &  1.09 &  19 &      13.62 &          6.39 &    4.5e-02 &       2.1e-02 &    5.2e-03 &       2.5e-03 &  1.634  \\ 
\label{Atab:yiping_data}
\end{longtable}


\begin{longtable}[!h]{cccc}
\caption{The 114 target systems, with data for our observations of each source. This includes the number of observations across the survey, and the average seeing and airmass for the observations of each source.}\label{tab:obs}
\\
\hline
\textbf{Survey Name}  & \textbf{Number of observations} & \textbf{Average Seeing (Arcseconds)} & \textbf{Average Airmass}\\
\hline
BELLSJ0830+5116 & 10 & 8.56 & 1.34\\
BELLSJ1159-0007 & 19 & 4.7 & 1.33\\
BELLSJ1221+3806 & 14 & 12.39 & 1.36\\
BELLSJ1318-0104 & 22 & 5.93 & 1.38\\
BELLSJ1337+3620 & 15 & 7.63 & 1.32\\
BELLSJ1349+3612 & 12 & 7.38 & 1.27\\
BELLSJ1542+1629 & 22 & 5.94 & 1.49\\
BELLSJ1601+2138 & 11 & 12.34 & 1.2\\
BELLSJ1631+1854 & 22 & 6.71 & 1.51\\
SDSSJ0029-0055 & 23 & 6.19 & 1.39\\
SDSSJ0037-0942 & 20 & 5.27 & 1.27\\
SDSSJ0044+0113 & 25 & 5.81 & 1.4\\
SDSSJ0109+1500 & 23 & 3.92 & 1.46\\
SDSSJ0143-1006 & 19 & 4.78 & 1.35\\
SDSSJ0157-0056 & 22 & 4.82 & 1.37\\
SDSSJ0159-0006 & 23 & 4.86 & 1.36\\
SDSSJ0216-0813 & 23 & 5.34 & 1.34\\
SDSSJ0252+0039 & 19 & 5.55 & 1.34\\
SDSSJ0324+0045 & 21 & 4.21 & 1.43\\
SDSSJ0324-0110 & 26 & 8.26 & 1.31\\
SDSSJ0330-0020 & 20 & 6.96 & 1.37\\
SDSSJ0405-0455 & 30 & 5.33 & 1.31\\
SDSSJ0728+3835 & 18 & 10.42 & 1.36\\
SDSSJ0737+3216 & 13 & 11.33 & 1.36\\
SDSSJ0753+3416 & 13 & 9.72 & 1.3\\
SDSSJ0754+1927 & 16 & 4.24 & 1.54\\
SDSSJ0757+1956 & 18 & 7.7 & 1.55\\
SDSSJ0822+1828 & 3 & 3.34 & 1.58\\
SDSSJ0822+2652 & 18 & 8.39 & 1.57\\
SDSSJ0847+2348 & 19 & 7.56 & 1.52\\
SDSSJ0851+0505 & 23 & 4.1 & 1.36\\
SDSSJ0903+4116 & 13 & 9.07 & 1.26\\
SDSSJ0912+0029 & 21 & 9.0 & 1.33\\
SDSSJ0920+3028 & 18 & 11.65 & 1.31\\
SDSSJ0935-0003 & 22 & 7.21 & 1.38\\
SDSSJ0936+0913 & 21 & 6.58 & 1.44\\
SDSSJ0946+1006 & 18 & 6.0 & 1.47\\
SDSSJ0955+0101 & 21 & 5.09 & 1.31\\
SDSSJ0955+3014 & 16 & 9.81 & 1.34\\
SDSSJ0956+5100 & 12 & 11.4 & 1.42\\
SDSSJ0959+0410 & 22 & 5.44 & 1.4\\
SDSSJ0959+4416 & 15 & 10.52 & 1.31\\
SDSSJ1010+3124 & 15 & 10.13 & 1.37\\
SDSSJ1016+3859 & 13 & 18.33 & 1.33\\
SDSSJ1020+1122 & 19 & 8.21 & 1.48\\
SDSSJ1031+3026 & 11 & 14.82 & 1.32\\
SDSSJ1032+5322 & 16 & 7.82 & 1.32\\
SDSSJ1040+3626 & 15 & 13.11 & 1.35\\
SDSSJ1041+0112 & 22 & 6.75 & 1.37\\
SDSSJ1048+1313 & 23 & 4.9 & 1.51\\
SDSSJ1051+4439 & 14 & 7.81 & 1.34\\
SDSSJ1056+4141 & 15 & 11.49 & 1.26\\
SDSSJ1100+5329 & 12 & 5.25 & 1.38\\
SDSSJ1101+1523 & 20 & 5.85 & 1.52\\
SDSSJ1103+5322 & 14 & 8.08 & 1.39\\
SDSSJ1106+5228 & 14 & 7.22 & 1.47\\
SDSSJ1112+0826 & 23 & 7.54 & 1.45\\
SDSSJ1127+2312 & 16 & 10.96 & 1.55\\
SDSSJ1137+1818 & 20 & 8.74 & 1.5\\
SDSSJ1142+1001 & 23 & 5.92 & 1.47\\
SDSSJ1142+2509 & 15 & 9.72 & 1.42\\
SDSSJ1143-0144 & 21 & 6.54 & 1.36\\
SDSSJ1144+0436 & 21 & 8.76 & 1.43\\
SDSSJ1153+4612 & 13 & 9.87 & 1.32\\
SDSSJ1204+0358 & 25 & 4.99 & 1.42\\
SDSSJ1205+4910 & 17 & 12.66 & 1.39\\
SDSSJ1213+2930 & 14 & 9.97 & 1.27\\
SDSSJ1213+6708 & 17 & 10.95 & 1.47\\
SDSSJ1250+0523 & 22 & 3.33 & 1.39\\
SDSSJ1251-0208 & 24 & 6.48 & 1.44\\
SDSSJ1301+0834 & 27 & 5.85 & 1.42\\
SDSSJ1330+1750 & 25 & 4.77 & 1.51\\
SDSSJ1416+5136 & 16 & 11.82 & 1.41\\
SDSSJ1420+6019 & 17 & 8.51 & 1.39\\
SDSSJ1430+4105 & 16 & 6.35 & 1.34\\
SDSSJ1432+6317 & 14 & 11.36 & 1.39\\
SDSSJ1433+2835 & 11 & 11.03 & 1.34\\
SDSSJ1436-0000 & 28 & 5.66 & 1.41\\
SDSSJ1443+0304 & 24 & 6.49 & 1.42\\
SDSSJ1451-0239 & 24 & 4.88 & 1.4\\
SDSSJ1525+3327 & 19 & 13.83 & 1.35\\
SDSSJ1538+5817 & 14 & 6.78 & 1.43\\
SDSSJ1541+3642 & 16 & 10.71 & 1.35\\
SDSSJ1543+2202 & 14 & 8.17 & 1.3\\
SDSSJ1550+2020 & 16 & 5.61 & 1.45\\
SDSSJ1553+3004 & 13 & 7.67 & 1.29\\
SDSSJ1607+2147 & 14 & 12.72 & 1.26\\
SDSSJ1621+3931 & 15 & 9.59 & 1.26\\
SDSSJ1627-0053 & 27 & 5.09 & 1.4\\
SDSSJ1630+4520 & 14 & 12.36 & 1.33\\
SDSSJ1633+1441 & 21 & 4.26 & 1.49\\
SDSSJ1636+4707 & 19 & 10.83 & 1.32\\
SDSSJ2238-0754 & 21 & 8.01 & 1.33\\
SDSSJ2300+0022 & 22 & 4.64 & 1.39\\
SDSSJ2303+1422 & 21 & 7.34 & 1.44\\
SDSSJ2309-0039 & 19 & 6.38 & 1.37\\
SDSSJ2321-0939 & 23 & 7.01 & 1.33\\
SWELLSJ0820+4847 & 11 & 13.08 & 1.39\\
SWELLSJ0822+1828 & 16 & 6.39 & 1.53\\
SWELLSJ0841+3824 & 15 & 15.29 & 1.25\\
SWELLSJ0915+4211 & 12 & 8.25 & 1.32\\
SWELLSJ0930+2855 & 13 & 9.45 & 1.28\\
SWELLSJ0955+0101 & 2 & 2.9 & 1.44\\
SWELLSJ1021+2028 & 17 & 7.25 & 1.55\\
SWELLSJ1029+0420 & 22 & 5.77 & 1.4\\
SWELLSJ1037+3517 & 11 & 5.65 & 1.3\\
SWELLSJ1111+2234 & 13 & 8.28 & 1.43\\
SWELLSJ1117+4704 & 13 & 8.74 & 1.3\\
SWELLSJ1135+3720 & 16 & 9.7 & 1.34\\
SWELLSJ1203+2535 & 16 & 16.7 & 1.27\\
SWELLSJ1313+0506 & 22 & 5.38 & 1.42\\
SWELLSJ1331+3638 & 14 & 11.96 & 1.39\\
SWELLSJ1703+2451 & 15 & 10.94 & 1.29\\
SWELLSJ2141\verb|-|0001 & 22 & 9.08 & 1.35\\
\label{Atab:obs_data}
\end{longtable}

\bsp	
\label{lastpage}
\end{document}